\documentclass[aps,showpacs,twocolumn,twoside,amsmath,amssymb,superscriptaddress,prc]{revtex4-2}
\usepackage{multirow}
\usepackage{tikz}
\usepackage{epstopdf}
\usepackage[percent]{overpic}  
\usepackage{scrextend}  
\usepackage{xcolor,xspace}
\usepackage[ulem=normalem]{changes}
\usepackage{hyperref}
\hypersetup{colorlinks=True,urlcolor=blue,linkcolor=blue,citecolor=blue,filecolor=black}

\colorlet{Changes@Color}{red}  
\usepackage{amsmath}
\usepackage{dcolumn,color,footnote,bm,braket}
\usepackage{url,longtable,tabularx}

\usepackage{fancybox}
\usetikzlibrary{matrix}

\newcommand\+{\dagger}

\begin{document}

\title{
Evolution of octupole deformation and collectivity in neutron-rich lanthanides
}

\author{K.~Nomura}
\email{knomura@phy.hr}
\affiliation{Department of Physics, Faculty of Science, University of Zagreb, HR-10000, Croatia}

\author{R.~Rodr\'iguez-Guzm\'an}
\affiliation{Physics Department, Kuwait University, 13060 Kuwait, Kuwait}

\author{L.~M.~Robledo}
\affiliation{Departamento de F\'\i sica Te\'orica and CIAFF, Universidad
Aut\'onoma de Madrid, E-28049 Madrid, Spain}

\affiliation{Center for Computational Simulation,
Universidad Polit\'ecnica de Madrid,
Campus de Montegancedo, Bohadilla del Monte, E-28660-Madrid, Spain
}

\author{J.~E.~Garc\'ia-Ramos}
\affiliation{Departamento de Ciencias Integradas y Centro de Estudios 
Avanzados en F\'isica, Matem\'atica y Computaci\'on, Universidad de Huelva, 
E-21071 Huelva, Spain}

\affiliation{Instituto Carlos I de F\'{\i}sica Te\'orica y Computacional,  
Universidad de Granada, Fuentenueva s/n, 18071 Granada, Spain}

\author{N.~C.~Hern\'andez}
\affiliation{Departamento de F\'isica Aplicada I, Escuela Polit\'ecnica Superior, Universidad de Sevilla, Seville, E-41011, Spain}

\date{\today}

\begin{abstract}
The onset of octupole deformation and its impact on related 
spectroscopic properties is studied in even-even neutron-rich 
lanthanide isotopes Xe, Ba, Ce, and Nd 
with neutron number $86\leqslant N\leqslant 94$. Microscopic 
input comes from the  
Hartree-Fock-Bogoliubov approximation with constrains
on the axially symmetric quadrupole and octupole operators using the Gogny-D1M interaction.
At the mean-field level, reflection asymmetric ground states
are predicted for isotopes with neutron number around $N=88$.
Spectroscopic properties are studied by  diagonalizing
the interacting boson model 
Hamiltonian, with the parameters obtained via 
the mapping of the mean-field potential energy surface onto the 
expectation value of the Hamiltonian in the 
$s$, $d$, and $f$ boson condensate state. The results 
obtained for low-energy  positive- and negative-parity 
excitation spectra as well as the  electric dipole, 
quadrupole, and octupole transition probabilities
indicate the onset of pronounced octupolarity
for  $Z\approx 56$ and 
$N\approx 88$ nuclei. 
\end{abstract}

\maketitle

\section{Introduction}

The ground state of most of medium-mass and 
heavy atomic nuclei is quadrupole deformed
and reflection symmetric. However, 
in specific regions of the nuclear chart, the 
spatial reflection symmetry is spontaneously
broken giving rise to  pear-like or octupole deformations 
\cite{butler1996,butler2016,butler2020b}. In particular, 
pronounced octupole correlations are expected 
in nuclei with neutron  
$N$ and/or proton $Z$ numbers close to the so-called ``octupole magic'' numbers 34, 56, 88, and 134 \cite{butler1996}. 
Typical examples are the 
light actinides near $Z=88$ and $N=134$ 
as well as  the neutron-rich lanthanides near $Z=56$ and $N=88$. 
Observables associated with static 
ground-state octupole deformation are low-lying negative-parity 
states and the  electric dipole ($E1$) and octupole ($E3$) 
transition strengths. Nowadays, nuclear octupolarity 
represents an active experimental research field. Fingerprints
of static ground-state octupole deformation have already been 
found in a number of nuclei
\cite{gaffney2013,bucher2016,bucher2017,butler2020a,chishti2020}. 

On the theoretical side, octupole deformation has
been studied using a large variety of approaches going from macroscopic 
models to very sophisticated microscopic calculations. We can mention calculations 
based on
macroscopic-microscopic models 
\cite{naza1984b,leander1985,moeller2008}, self-consistent 
mean-field (SCMF) approaches with and without symmetry restoration 
\cite{MARCOS1983,BONCHE1986,BONCHE1991,heenen1994,ROBLEDO1987,ROBLEDO1988,EGIDO1990,EGIDO1991,EGIDO1992,GARROTE1998,GARROTE1999,long2004,robledo2010,robledo2011,erler2012,robledo2012,rayner2012,robledo2013,robledo2015,bernard2016,agbemava2016,agbemava2017,xu2017,xia2017,ebata2017,rayner2020,cao2020,rayner2020oct,rayner2021,nomura2021qoch}, 
interacting boson models (IBM)
\cite{engel1985,engel1987,kusnezov1988,yoshinaga1993,SMIRNOVA2000,zamfir2001,zamfir2003,pietralla2003,nomura2013oct,nomura2014,nomura2015,nomura2020oct,nomura2021oct,vallejos2021}, 
shell model \cite{yoshinaga2018,yoshinaga2021}, 
geometrical collective models 
\cite{bonatsos2005,lenis2006,bizzeti2013}, 
and cluster models \cite{shneidman2002,shneidman2003}. 
Among the SCMF approaches it is worth noticing the 
recent calculations within the framework of the full  
symmetry-restored (angular momentum, particle number and parity) generator coordinate method (GCM) \cite{RS} based on 
the  Gogny \cite{bernard2016,bucher2017,lica2018} and covariant 
\cite{fu2018} EDFs performed to analyze octupole correlations in the low-lying states of nuclei around 
$^{144}$Ba. However, for heavy nuclear systems, full 
symmetry-projected GCM calculations are quite time consuming. 
Therefore, alternative approaches such as the full axial quadrupole-octupole 
two-dimensional GCM \cite{robledo2013} or the 
collective Hamiltonian, which is an approximation to the GCM based on the Gaussian overlap
approximation, are often employed 
\cite{ROBLEDO1988,xia2017,nomura2021qoch}.

In this work, we consider the spectroscopy of the quadrupole and 
octupole collective states in lanthanide nuclei with  proton and 
neutron numbers close to the ``octupole magic'' numbers 56 and 88, 
respectively. Similarly to the  light Ra and Th isotopes, the 
above-mentioned neutron-rich lanthanide nuclei are expected to exhibit 
enhanced octupolarity. The study is motivated by recent Coulomb 
excitation experiments performed at the CARIBU facility of Argonne 
National Laboratory which reveal a substantially large $B(E3)$ 
transition probability in  $^{144}$Ba \cite{bucher2016} and $^{146}$Ba 
\cite{bucher2017}, typical of a well octupole deformed nucleus.

For the evaluation of the spectroscopic observables we use the (mapped) 
IBM framework based on  input provided by a microscopic  energy density 
functional (EDF). First, for each of the studied nuclei, a potential 
energy surface (PES) is obtained as a function of the 
(axially-symmetric) quadrupole $\beta_{2}$ and octupole $\beta_{3}$ 
deformation parameters. To obtain such a PES, we rely on the 
constrained SCMF approximation based on the parametrization D1M of the 
Gogny interaction \cite{D1M}. Second, the PES (hereafter denoted by 
SCMF-PES) is mapped onto the corresponding expectation value of the IBM 
Hamiltonian in the condensate state consisting of the monopole $s$, 
quadrupole $d$, and octupole $f$ bosons, with spin and parity $0^{+}$, 
$2^{+}$, and $3^{-}$, respectively \cite{engel1985,engel1987,IBM}. The 
strength parameters of the $sdf$-IBM Hamiltonian are determined  by the 
mapping procedure. The diagonalization of the $sdf$-IBM Hamiltonian 
then yields positive- and negative-parity excitation spectra as well as 
electromagnetic transition probabilities. The method has already been 
used to study quadrupole-octupole shape phase transitions in 
reflection-asymmetric Ra, Th, Sm, and Ba nuclei 
\cite{nomura2013oct,nomura2014} using the relativistic DD-PC1 EDF 
\cite{DDPC1} as microscopic input, and the spectroscopic properties of 
deformed rare-earth Sm and Gd nuclei \cite{nomura2015} using the 
Gogny-D1M  EDF \cite{D1M}.

The mapped IBM framework, based on the Gogny-D1M 
Hartree-Fock-Bogoliubov (HFB) approach, has been recently applied to 
carry out spectroscopic calculations for even-even Ra, Th, U, Pu, Cm, 
and Cf isotopes \cite{nomura2020oct,nomura2021oct}. Those studies point 
towards the onset of stable octupole deformation around $N=134$ as well 
as the development of octupole softness  from $N\approx 138$ on. Within 
this context, and motivated by the renewed experimental interest in 
octupole correlations, it is meaningful and timely to extend the 
calculations of Refs.~\cite{nomura2020oct,nomura2021oct} to the 
lanthanide region to obtain  updated theoretical predictions on 
octupole-related spectroscopic properties. To this end, we have 
considered in this paper the even-even neutron-rich Xe, Ba, Ce, and Nd 
isotopes with neutron numbers $86\leqslant N\leqslant 94$. We study the 
evolution of the quadrupole-octupole coupling in those nuclei as well 
as the appearance of stable octupole deformation around $N=88$.

The paper is organized as follows. The SCMF-to-IBM mapping procedure is 
outlined in Sec.~\ref{sec:method}. The results of the calculations are 
discussed in Sec.~\ref{sec:results}. In this section, attention is paid 
to the Gogny-D1M  and mapped IBM PESs, low-lying positive and negative 
parity states as well as to the $B(E1)$, $B(E2)$, and $B(E3)$ 
transition strengths in the studied Xe, Ba, Ce, and Nd nuclei. 
Section~\ref{sec:summary} is devoted to the concluding remarks.

\section{Theoretical method \label{sec:method}}

To obtain the SCMF-PES, the HFB equation has been
solved with constrains on the axially symmetric quadrupole
$\hat{Q}_{20}$ and octupole $\hat{Q}_{30}$ operators
\cite{rayner2012,rayner2020oct}. 
The mean values $\langle \Phi_\mathrm{HFB} |\hat{Q}_{20}| \Phi_\mathrm{HFB} \rangle = Q_{20}$
and $\langle \Phi_\mathrm{HFB} |\hat{Q}_{30}| \Phi_\mathrm{HFB} \rangle
= Q_{30}$ define the quadrupole and octupole deformation parameters
$\beta_{\lambda}$ ($\lambda=2,3$), i.e.,  $\beta_{\lambda} =
\sqrt{4 \pi (2\lambda +1)}Q_{\lambda 0}/(3 R_{0}^{\lambda} A)$,  
with $R_0=1.2 A^{1/3}$ fm. The constrained  calculations 
provide a set of HFB states $| \Phi_\mathrm{HFB} (\beta_{2},\beta_{3})\rangle$ 
labeled by their  
static deformation parameters $\beta_{2}$ and $\beta_{3}$.
The energies $E_\mathrm{HFB}(\beta_{2},\beta_{3})$ associated
with those Gogny-HFB states define the 
SCMF-PESs. Note, that the HFB energies satisfy the 
property 
$E_\mathrm{HFB}(\beta_{2},\beta_{3}) = E_\mathrm{HFB}(\beta_{2},-\beta_{3})$. 
Therefore, only positive $\beta_{3}$ values
are considered.

The Gogny-D1M  SCMF-PES is subsequently 
mapped onto the $sdf$-IBM Hamiltonian via the  
procedure briefly described below. The building blocks 
of the IBM, i.e., the $s$, $d$, and $f$ bosons,  
represent, from a microscopic point of view 
\cite{OAI,IBM}, collective monopole, quadrupole, and  
octupole pairs of valence nucleons, respectively. 
The total number of bosons $n=n_{s}+n_{d}+n_{f}$ is equal 
to half the number of valence nucleons, and is conserved for 
a given nucleus. 
In contrast to the majority of previous $sdf$-IBM phenomenology, 
which assumed the number of $f$ bosons to be 
$n_{f}\leqslant1$ or $n_{f}\leqslant3$, 
here $n_{f}$ is allowed to take any value between zero and $n$.

In principle, one could 
make distinction between proton and neutron degrees of freedom 
within the framework of the proton-neutron IBM (IBM-2) 
\cite{OAI,mizusaki1996}. 
The IBM-2 framework, however, involves a large number 
of model parameters, especially when it includes the $f$ boson 
degrees of freedom. 
It is perhaps for this reason 
that the IBM-2 has rarely been considered in 
dealing with the quadrupole and octupole collective states, 
except for a few instances 
\cite{yoshinaga1993,SMIRNOVA2000,pietralla2003,vallejos2021}. 
On the other hand, the simpler $sdf$-IBM-1 framework, which does 
not distinguish between proton and neutron bosons, 
has been successfully applied to a number of phenomenological studies. 
Within this context, and in order to
keep our description as simple as possible, we resort in this paper to the $sdf$-IBM-1 as in our 
previous studies of octupole correlations 
\cite{nomura2020oct,nomura2021oct}.

%
%
\begin{figure*}[htb!]
\begin{center}
\includegraphics[width=\linewidth]{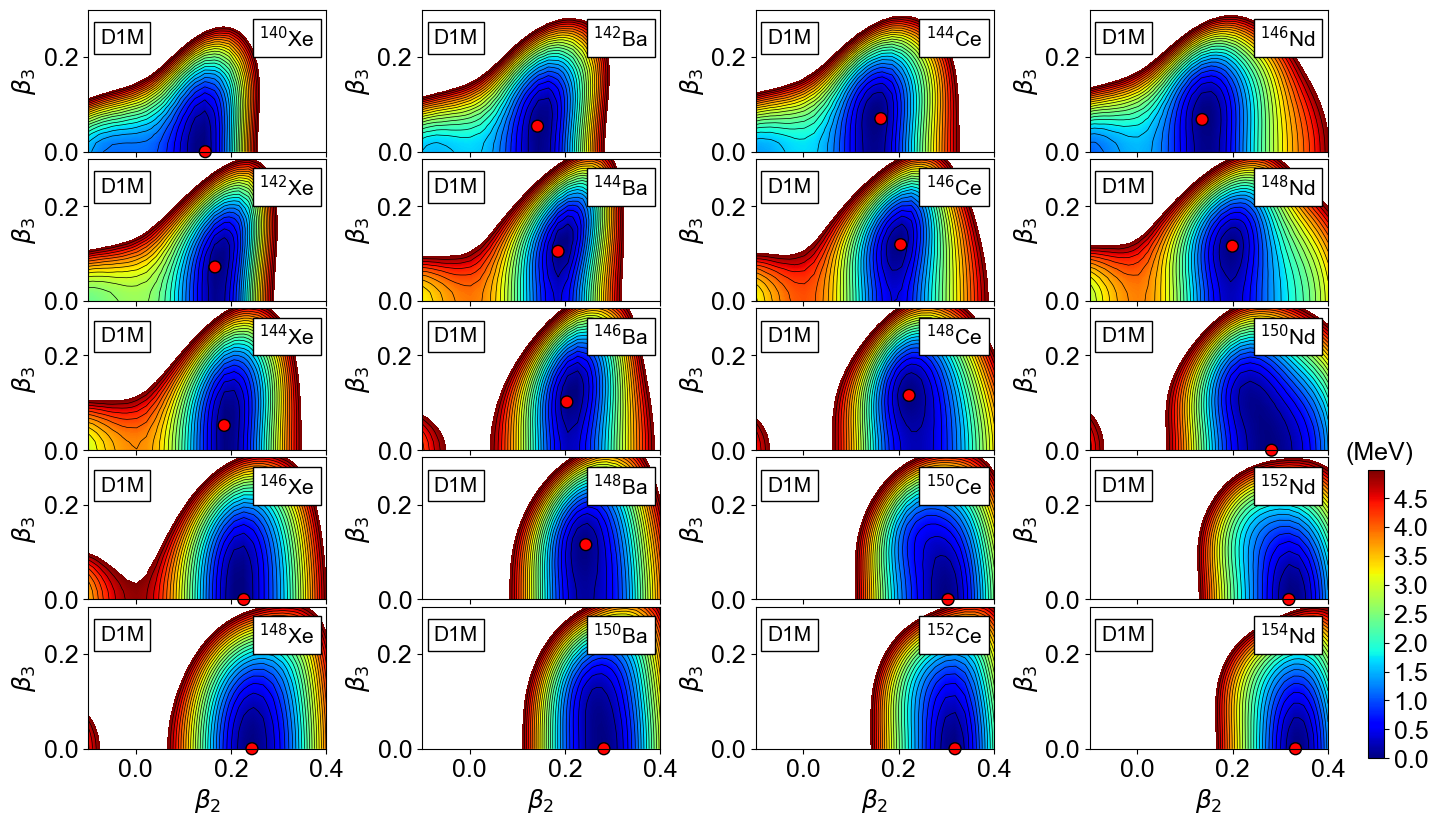}
\caption{The SCMF-PESs obtained for 
$^{140-148}$Xe, $^{142-150}$Ba, $^{144-152}$Ce, 
and $^{146-154}$Nd are plotted as functions of the 
quadrupole $\beta_{2}$ and octupole $\beta_{3}$ deformation 
parameters. The color code indicates the total HFB energies (in MeV) 
plotted up to 5 MeV with respect to the global minimum. The 
energy
 difference between neighboring contours is 0.2 MeV. For each
 nucleus, the global minimum is
 indicated by a red solid circle. Results have been obtained with the 
 Gogny-D1M EDF.} 
\label{fig:pesdft}
\end{center}
\end{figure*}
%
%
\begin{figure*}[htb!]
\begin{center}
\includegraphics[width=\linewidth]{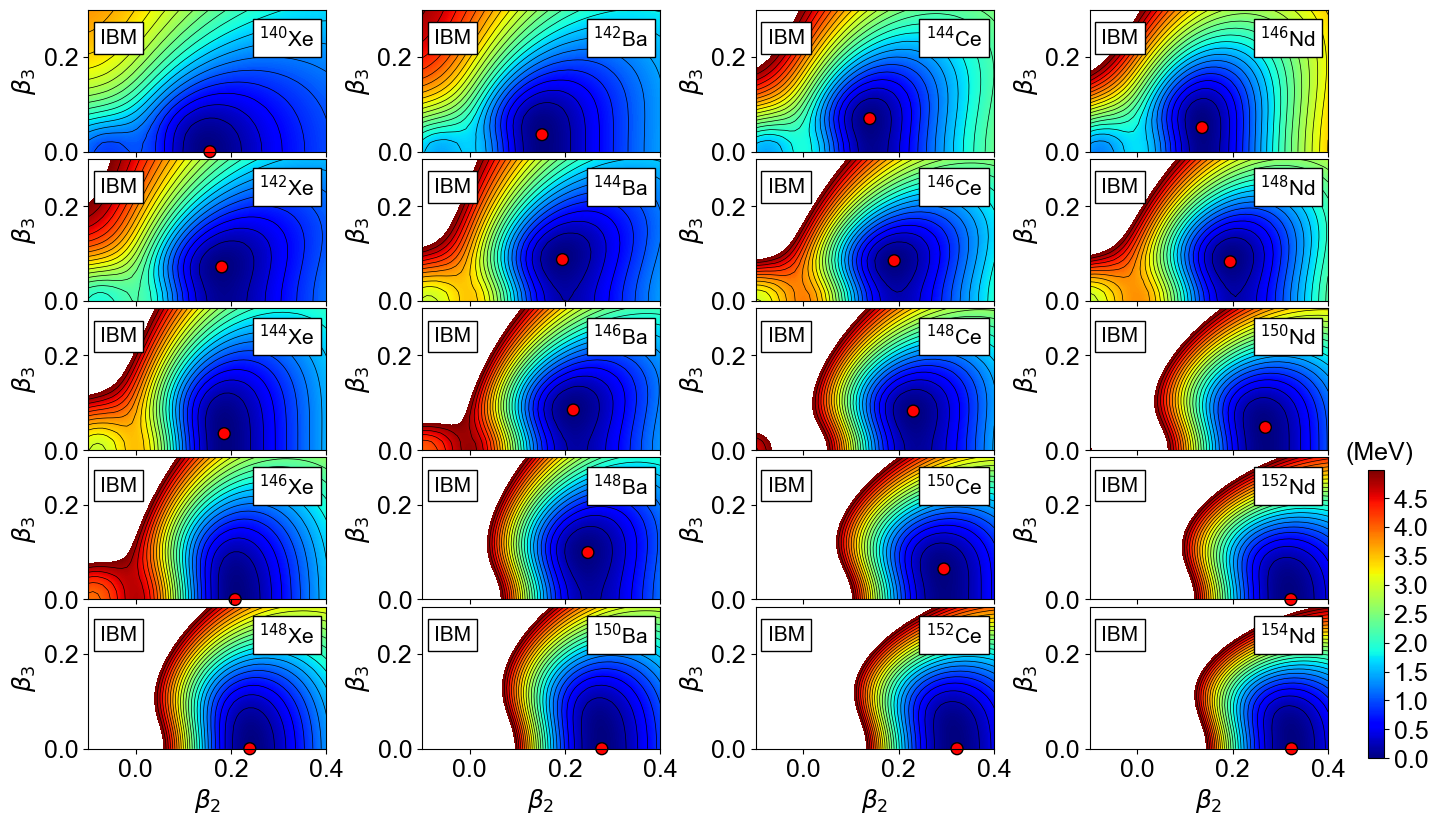}
\caption{The same as in 
Fig.~\ref{fig:pesdft}, but for the mapped IBM-PESs.} 
\label{fig:pesibm}
\end{center}
\end{figure*}

We have employed the same $sdf$-IBM Hamiltonian 
as in our previous studies for actinide nuclei 
\cite{nomura2020oct,nomura2021oct}:
\begin{align}
\label{eq:ham}
\hat H= 
\epsilon_d\hat n_{d} + \epsilon_{f}\hat{n}_{f} 
+ \kappa_{2}\hat{Q}_{2}\cdot\hat{Q}_{2} + \rho\hat{L}\cdot\hat{L} 
+ \kappa_{3}\hat{Q}_{3}\cdot\hat{Q}_{3}. 
\end{align}
The first (second) term represents the number operator for
the $d$ ($f$) bosons with $\epsilon_{d}$ ($\epsilon_{f}$) being the single
$d$ ($f$) boson energy relative to the $s$ boson one. 
The third, fourth  and fifth terms represent the quadrupole-quadrupole
interaction, the rotational term, and the octupole-octupole 
interaction, respectively. 
The quadrupole $\hat{Q}_{2}$, the angular momentum $\hat{L}$, and the
octupole $\hat{Q}_{3}$
operators read 
\begin{subequations}
 \begin{align}
\label{eq:q2}
& \hat Q_{2}=s^{\dagger}\tilde d+d^{\dagger}\tilde s+\chi_{d}(d^{\dagger}\tilde
  d)^{(2)}+\chi_{f}(f^{\dagger}\tilde f)^{(2)} \\
\label{eq:l}
& \hat{L}=
\sqrt{10}(d^{\dagger}\tilde{d})^{(1)}
+\sqrt{28}(f^\+\tilde{f})^{(1)} \\
\label{eq:q3}
&\hat{Q}_{3}=
s^{\dagger}\tilde{f}+f^{\dagger}\tilde{s}
+\chi_{3}(d^{\dagger}\tilde{f}
+f^{\dagger}\tilde{d})^{(3)}. 
\end{align}
\end{subequations}
Note that the term proportional to
$(d^{\+}\tilde{d})^{(1)}\cdot (f^{\+}\tilde{f})^{(1)}$ in the
$\hat{L}\cdot\hat{L}$ term has been neglected \cite{nomura2020oct}. 
The parameters 
$\epsilon_d$, $\epsilon_f$, $\kappa_{2}$, $\rho$, $\chi_{d}$, $\chi_{f}$, 
$\kappa_{3}$, and $\chi_{3}$ of the $sdf$-IBM Hamiltonian 
are determined, for each nucleus, in such a way  
\cite{nomura2015,nomura2020oct} that 
the expectation value of the $sdf$-IBM Hamiltonian 
in the boson condensate state (denoted by IBM-PES),
$E_\mathrm{IBM}(\beta_2,\beta_3)=
\bra{\phi(\beta_{2},\beta_{3})}\hat{H}\ket{\phi(\beta_{2},\beta_{3})}$, 
reproduces the SCMF-PES $E_\mathrm{HFB}(\beta_2,\beta_3)$ in
the neighborhood of the global minimum. 
The boson condensate state is given by 
\cite{ginocchio1980}: 
\begin{subequations}
 \label{eq:coherent}
 \begin{align}
|\phi(\beta_{2},\beta_{3})\rangle=
(n!)^{-1/2}(b_{c}^{\+})^{n}
\ket{0}
 \end{align}
with
 \begin{align}
b_{c}^{\+}=(1+{\bar\beta_{2}}^{2}+{\bar\beta_{3}}^{2})^{-1/2}
(s^\+ +\bar\beta_{2}d_0^\+ + \bar\beta_{3}f_{0}^{\+}),
\end{align}
\end{subequations}
where $\ket{0}$ denotes the boson vacuum, or inert core. 
In the present work, the doubly-magic nucleus 
$^{132}$Sn is taken as the inert core, hence 
$n=(A-132)/2$ for a nucleus with mass number $A$. 
The amplitudes $\bar\beta_{2}$ and $\bar\beta_{3}$ entering the 
definition of the boson condensate wave function are assumed to be 
proportional to the deformation parameters $\beta_{2}$ and $\beta_{3}$ 
of the fermionic space, $\bar\beta_{2}=C_2\beta_2$ and
$\bar\beta_{3}=C_3\beta_3$ \cite{ginocchio1980,nomura2014,nomura2015}, 
with dimensionless proportionality constants $C_2$ and $C_3$. 
The reason for the scaling is that the present version of the IBM is built on a restricted model 
space of valence nucleons in one major shell, while the SCMF model 
is defined on the entire fermion Hilbert space. 
The difference between the fermionic and bosonic model spaces is 
effectively accounted for by the constants $C_{2}$ and $C_{3}$. 
Their values are also determined by the
mapping procedure so that the location of the global minimum 
in the SCMF-PES 
is reproduced. 
The parameter $\rho$ is fixed separately 
\cite{nomura2011rot} by equating the cranking moment of inertia 
obtained in the intrinsic frame of the IBM 
\cite{schaaser1986}
at the equilibrium minimum to the corresponding 
Thouless-Valatin value \cite{TV} 
computed by the Gogny-HFB method.  
A more detailed description of the whole procedure 
can be found in
Ref.~\cite{nomura2020oct}. 
The analytical form of the IBM-PES
$E_\mathrm{IBM}(\beta_2,\beta_3)$ is given
in Ref.~\cite{nomura2015}. 
For the numerical diagonalization of the 
mapped Hamiltonian $\hat{H}$ (\ref{eq:ham}), 
the computer code \textsc{arbmodel} \cite{arbmodel} 
has been  used.

%
%
\begin{figure}[htb!]
\begin{center}
\includegraphics[width=\linewidth]{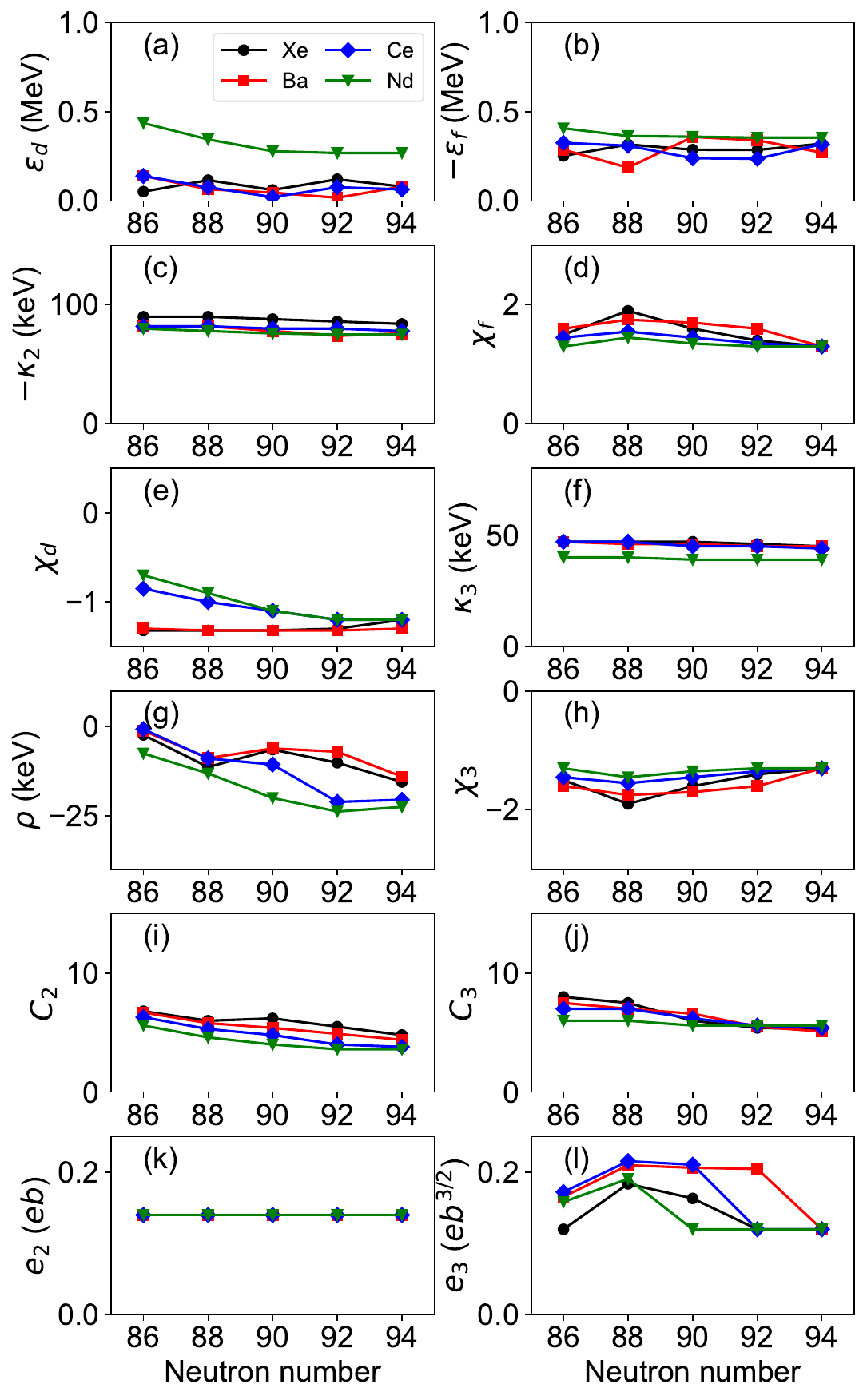}
\caption{The strength parameters (a)
$\epsilon_d$, (b) $\epsilon_f$, (c) $\kappa_2$, 
(d) $\chi_{f}$, (e) $\chi_{d}$, (f) $\kappa_{3}$, 
(g) $\rho$, and (h) $\chi_{3}$ of the $sdf$-IBM 
Hamiltonian (\ref{eq:ham}), and the coefficients 
(i) $C_{2}$ and (j) $C_{3}$ 
are plotted as functions of the neutron number 
for the studied isotopic chains.
The boson effective charges for the quadrupole 
$e_{2}$ and octupole $e_{3}$
transitions are also plotted in panels (k) and (l), respectively.}
\label{fig:parameter}
\end{center}
\end{figure}


\section{Results and discussions\label{sec:results}}


In this section we discuss the results of the calculations.
Attention is paid to the SCMF- and IBM-PESs in Sec.~\ref{sec:pes}, 
the evolution of the low-energy excitation spectra 
in Sec.~\ref{sec:spectra}, alternating-parity structures 
in Sec.~\ref{sec:alt} and  transition rates in Sec.~\ref{sec:trans}.
The detailed spectroscopy of selected nuclei is considered in 
Sec.~\ref{sec:detail}.

\subsection{Potential energy surfaces\label{sec:pes}}

The $(\beta_{2}$, $\beta_{3})$  Gogny-HFB SCMF-PESs obtained 
for the considered 
Xe, Ba, Ce, and Nd nuclei are depicted 
in Fig.~\ref{fig:pesdft}.
In a number of nuclei close to the neutron 
number $N=88$ a minimum with 
$\beta_{3}\approx 0.1$ is obtained. 
The most prominent example 
is the Ba chain where four of the studied
isotopes exhibit a (static)  octupole-deformed ground state.
These SCMF results are consistent with the 
empirical fact that octupole correlations 
are enhanced near  $Z=56$ and  
$N=88$. For heavier nuclei, the ground state 
is reflection symmetric 
while the potential is still rather soft along the 
$\beta_{3}$-direction. Note, that this octupole 
softness, characteristic of an octupole vibrational
regime, indicates that octupole correlations
are still important even though the global  
minimum of the PESs occur at $\beta_{3}= 0$.
On the other hand, the  $\beta_{2}$ value at the minimum 
increases with neutron number. 

In Fig.~\ref{fig:pesibm} we have plotted  the 
corresponding IBM-PESs. 
Comparing with Fig.~\ref{fig:pesdft} we conclude that the 
essential features of the SCMF-PESs in the vicinity 
of the global minimum are reproduced by the 
IBM-PESs. 
However, some differences are observed: for $^{150}$Ce and $^{150}$Nd
the IBM-PESs exhibit global minima at non-zero $\beta_{3}$ 
deformation, whereas the SCMF-PESs for these nuclei 
show reflection symmetric ($\beta_{3}=0$) minima.
A closer look reveals that those discrepancies are rather insignificant 
and they should have little impact on the spectroscopic properties: the
depth of the octupole deformed IBM-PES minima do not exceed 20 keV and the
SCMF-PES are  are also considerably soft 
along the $\beta_{3}$ direction. 
For lighter isotopes, the IBM-PESs are much flatter than 
the SCMF-PESs. This is a feature already observed 
in previous studies using the SCMF-to-IBM mapping procedure 
\cite{nomura2020oct,nomura2021oct}. 
It is another consequence of the mapped IBM being built on 
a restricted model (valence) space while the 
SCMF model considers all the nucleons.

The parameters of the mapped IBM Hamiltonian
are shown in Figs.~\ref{fig:parameter}(a) to 
\ref{fig:parameter}(h)
while the proportionality constants 
$C_{\lambda}$ are depicted 
in Figs.~\ref{fig:parameter}(i) and \ref{fig:parameter}(j).
Most of the Hamiltonian parameters exhibit 
a gradual change as functions of the neutron number
and their systematic trend is similar 
in neighboring isotopic chains. The stability of those 
parameters as functions of the  nucleon 
number is particularly important and satisfying, as it reveals 
the robustness of the mapping procedure in 
the considered mass region.

%
%
\begin{figure}[htb!]
\begin{center}
\includegraphics[width=\linewidth]{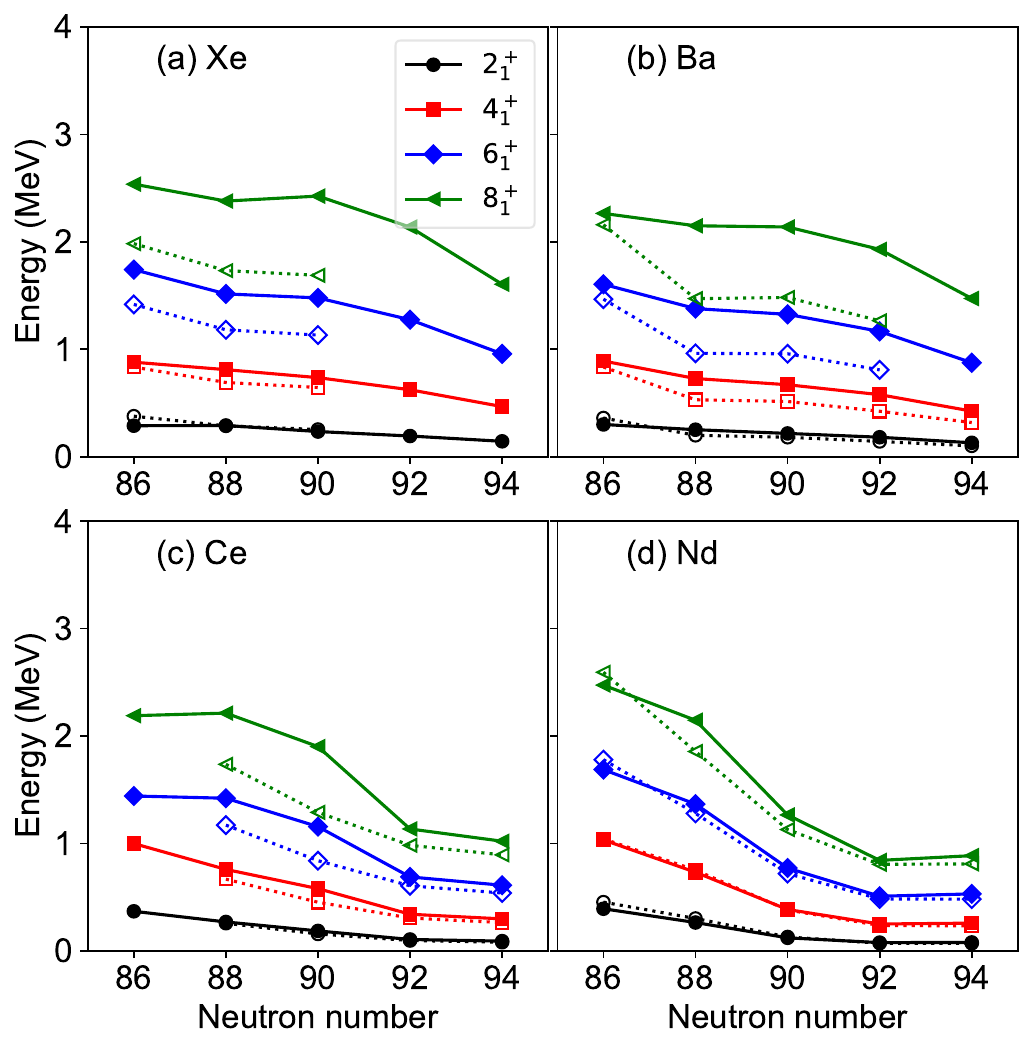}
\caption{The low-energy excitation spectra of 
positive-parity even-spin
 yrast states in $^{140-148}$Xe, $^{142-150}$Ba, 
$^{144-152}$Ce and
 $^{146-154}$Nd, computed by  diagonalizing
the mapped $sdf$-IBM Hamiltonian (\ref{eq:ham}), are
shown as functions of the neutron number. 
Experimental data have been taken from Ref.~\cite{data}.} 
\label{fig:level-pos}
\end{center}
\end{figure}
%
%
\begin{figure}[htb!]
\begin{center}
\includegraphics[width=\linewidth]{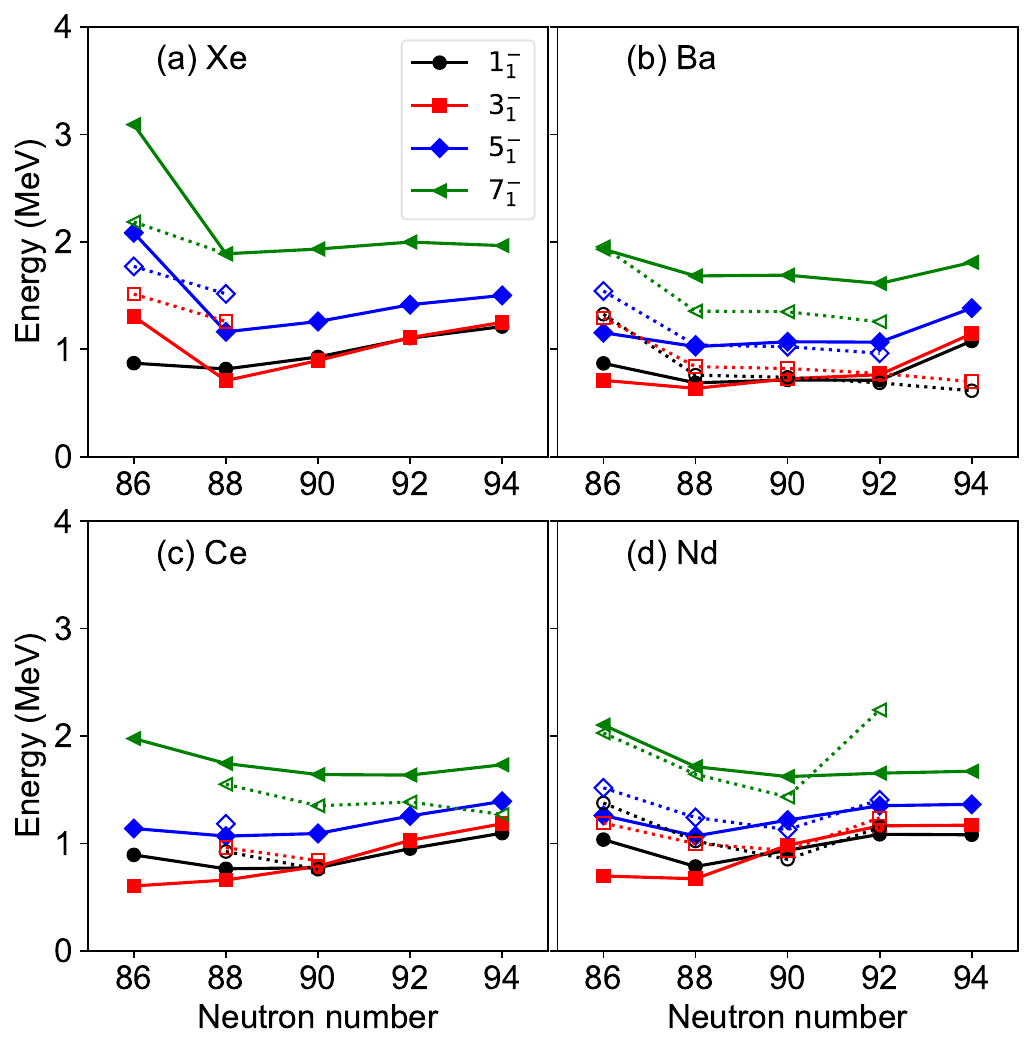}
\caption{The same as in Fig.~\ref{fig:level-pos} 
but for odd-spin negative-parity yrast states.} 
\label{fig:level-neg}
\end{center}
\end{figure}
%
%
\begin{figure}[htb!]
\begin{center}
\includegraphics[width=\linewidth]{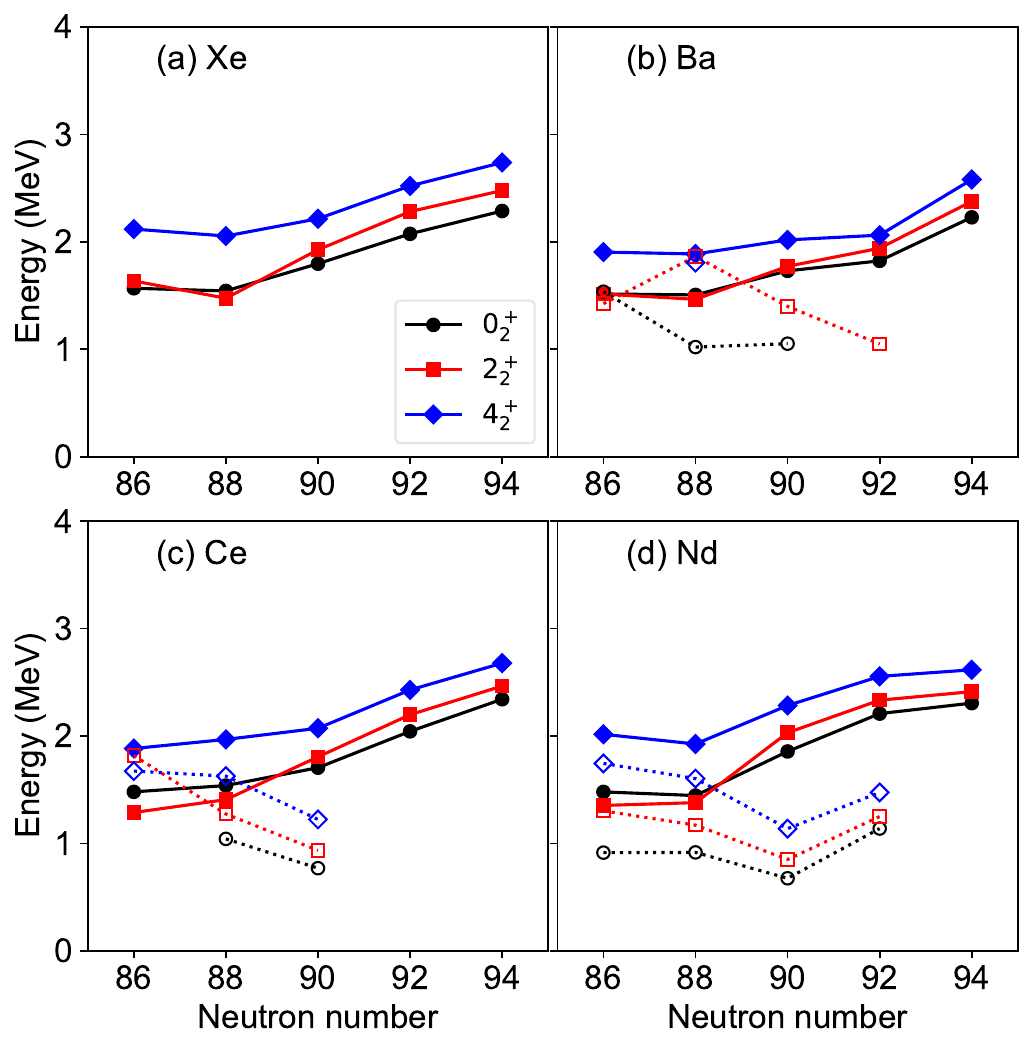}
\caption{The same as in Fig.~\ref{fig:level-pos} 
but for the $0^{+}_{2}$, $2^{+}_{2}$, and $4^{+}_{2}$ states. 
The experimental data are taken from Refs.~\cite{data,zhu2020}.} 
\label{fig:level-beta}
\end{center}
\end{figure}
%
%
\begin{figure}[htb!]
\begin{center}
\includegraphics[width=\linewidth]{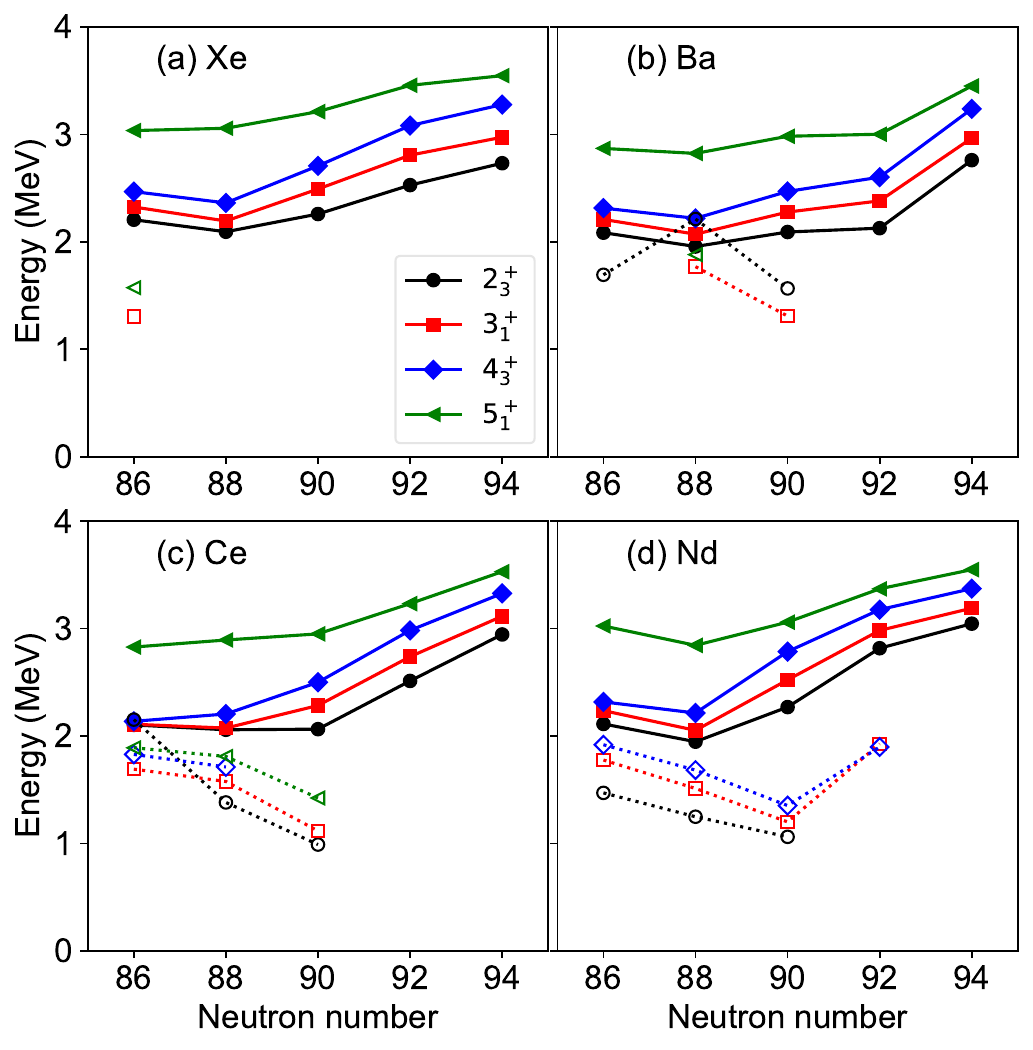}
\caption{The same as in Fig.~\ref{fig:level-pos} 
but for the $2^{+}_{3}$, $3^{+}_{1}$, $4^{+}_{3}$, 
and $5^{+}_{1}$ states.
The experimental data are taken from Refs.~\cite{data,zhu2020}.} 
\label{fig:level-gamma}
\end{center}
\end{figure}
%
%
\begin{figure}[htb!]
\begin{center}
\includegraphics[width=\linewidth]{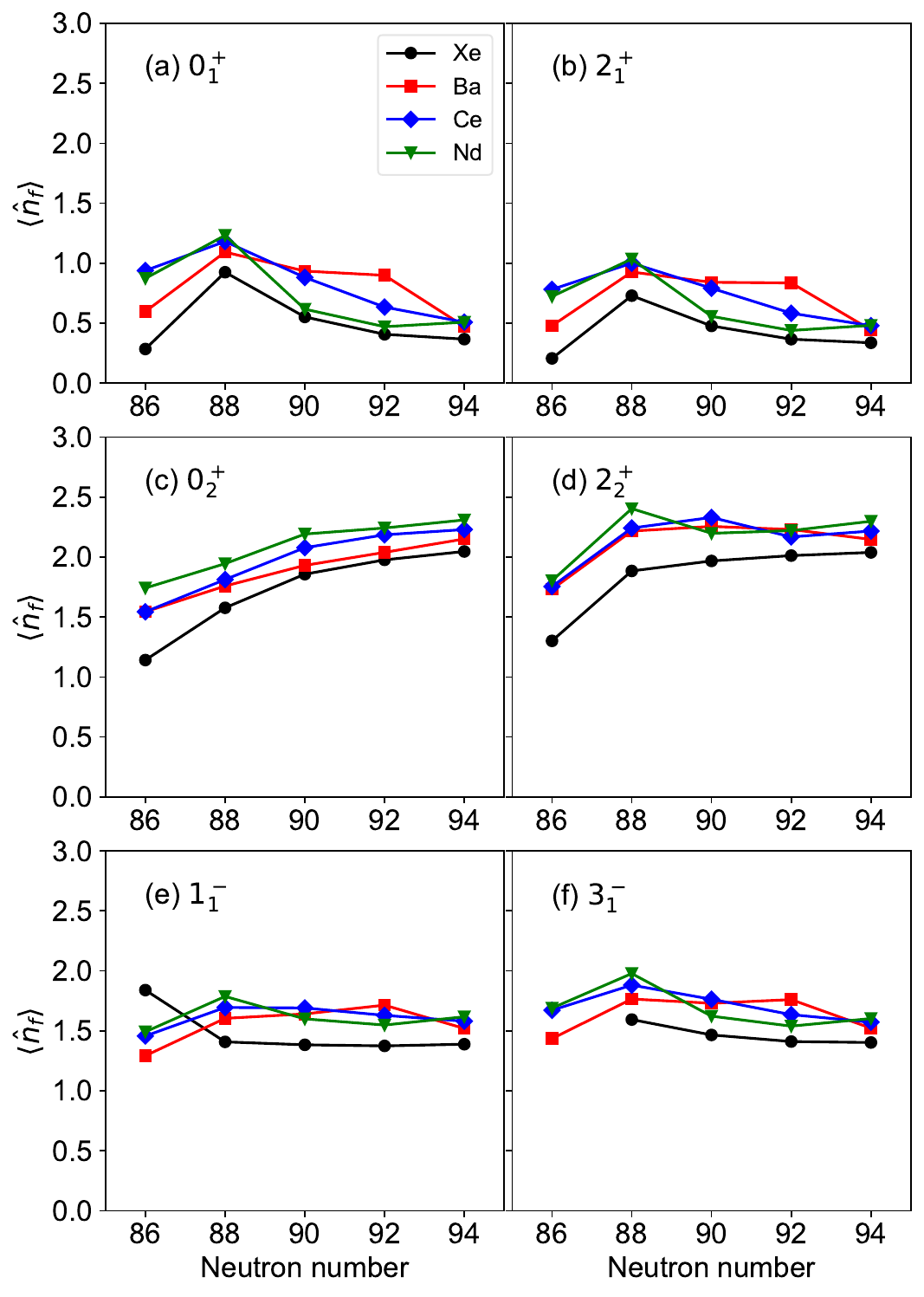}
\caption{The expectation values 
of the $f$-boson number operator 
$\braket{\hat{n_{f}}}$ in the IBM wave functions
corresponding to the  
states (a) $0^+_1$, (b) $2^{+}_{1}$, 
(c) $0^{+}_{2}$, (d) $2^{+}_{2}$, 
(e) $1^{-}_{1}$, and (f) $3^{-}_{1}$ 
are plotted  as functions of the neutron number.} 
\label{fig:nf}
\end{center}
\end{figure}

\subsection{Evolution of low-energy excitation spectra\label{sec:spectra}}

The excitation spectra corresponding to the 
positive-parity even-spin and negative-parity 
odd-spin yrast states obtained in our calculations 
for Xe, Ba, Ce, and Nd nuclei are plotted in 
Figs.~\ref{fig:level-pos} 
and \ref{fig:level-neg}, respectively. The
available experimental data taken 
from the NNDC database \cite{data}
are shown in the same figures.

As can be seen from Fig.~\ref{fig:level-pos}, both 
the theoretical and experimental positive-parity 
yrast states exhibit a monotonic decrease 
as functions of the neutron number. This is a 
typical feature of a near-spherical-to-deformed shape 
transition. The impact of the shape phase transitions in the 
positive-parity states in the neutron-rich Ba region 
were also considered in empirical studies 
(e.g., Refs.~\cite{sugawara2007,gupta2015,lee2018}). 
In the case of the Ce and Nd isotopes, the 
calculations predict a pronounced  structural change 
between 
$N=90$ and 92. The  positive-parity 
ground-state band obtained with the mapped IBM
model exhibits a qualitatively similar pattern when compared with
the experimental one. At the quantitative level  most 
of the studied nuclei with $N<92$ show a considerably  
stretched theoretical ground-state band implying that the moment of inertia is too small 
as compared with the experimental one. 
This is probably a consequence of the too large 
quadrupole-quadrupole interaction strength $\kappa_{2}$ 
determined by the mapping procedure. 
The value of the derived parameter, in turn, reflects 
the topology of the Gogny-D1M SCMF-PES which 
exhibits a pronounced minimum with a non-zero octupole 
deformation $\beta_{3}$ (see, Fig.~\ref{fig:pesdft}). 
To reproduce such a topology 
of the SCMF-PES in the IBM energy surface a large value of the strength parameter 
$\kappa_{2}$ is required. 
Another likely explanation is that, especially for lighter 
nuclei near the shell closure, 
the number of bosons is not large enough to 
describe satisfactorily the excitation energies 
with spin $I\geqslant 6^{+}$.

A characteristic signature of octupole collectivity 
is the lowering of the low-lying negative-parity 
states with respect to the positive-parity 
ground-state band. As can be seen from
Fig.~\ref{fig:level-neg}, for each of the studied 
isotopic chains, such a 
pattern is observed  
in both the theoretical 
and experimental negative-parity levels. The 
excitation energies of the 
predicted negative-parity bands decrease 
toward $N\approx 88$. For 
 most of the studied isotopic chains 
the $3^{-}$ excitation energy reaches its 
minimum value around this neutron number.
For $N>88$ the 
excitation energies of the 
negative-parity levels 
gradually increase. This reflects 
the fact that, at the HFB level, 
the octupole minimum becomes less 
pronounced as $N$ increases.
However, the negative-parity excitation 
energies remain rather constant
for the Ba isotopes up to $N=92$.
At the HFB level, an 
octupole-deformed ground state is indeed 
found in Ba isotopes with neutron number between 86 and 92 
(cf. Figs.~\ref{fig:pesdft} and \ref{fig:pesibm}). 
Similar results are obtained for Ce isotopes.
In the case of Xe and Nd isotopes, an approximate 
parabolic systematic is observed in the 
predicted levels 
around $N=88$ and 90. The systematic of the negative-parity 
states already discussed, suggests that in the 
neutron-rich lanthanide region, octupole
collectivity evolves moderately.
This is at variance with the structural change observed
in light actinides, especially Ra and Th isotopes. 
In that case the parabolic dependence on neutron 
number of the negative-parity band energy is stronger \cite{nomura2021oct},
with the lowest energy 
at $N\approx 134$. The 
negative-parity bands predicted
for Ba and Xe isotopes are 
systematically stretched as compared with the 
experimental ones. The reason is the same 
as in the case of the positive-parity states 
in Fig.~\ref{fig:level-pos}.

Figure~\ref{fig:level-beta} displays the
excitation energies of the non-yrast states 
$0^{+}_{2}$, $2^{+}_{2}$, and $4^{+}_{2}$. 
For nuclei with $N\geqslant90$  those states 
appear to form a quasi-$\beta$ band with the $0^{+}_{2}$  state as the 
bandhead. 
For the transitional nuclei with $N<90$,  
Fig.~\ref{fig:level-beta} shows that the computed energy levels 
display a more irregular pattern, with an  inversion of 
the position of the $0^{+}_{2}$ and $2^{+}_{2}$ energy levels. 
In the transitional region, 
the SCMF-PESs are rather soft both along the $\beta_{2}$ 
and $\beta_{3}$ deformations, indicating considerable shape mixing. 
As a consequence, the level repulsion among low-spin states 
is so strong as to explain the irregular
band structure.
In addition, as one can see most noticeably in the 
strongly quadrupole deformed Ce and Nd isotopes in 
Figs.~\ref{fig:level-beta}(c) and \ref{fig:level-beta}(d), 
respectively, the calculated energy levels are considerably 
higher than the experimental ones. 
This is a common feature in  mapped 
IBM studies, that is mainly due to the unexpectedly large 
$\kappa_{2}$ value.

Figure~\ref{fig:level-gamma} depicts the excitation spectra 
for another set of non-yrast states, 
$2^{+}_{3}$, $3^{+}_{1}$, $4^{+}_{3}$, and $5^{+}_{1}$. 
The present calculation suggests that in most of the considered nuclei 
these states are members of quasi-$\gamma$ bands. 
In fact, as can be seen in Fig.~\ref{fig:level-gamma}, 
the energy levels look more or less harmonic. Only the $5^{+}_{1}$ 
level, especially 
for lighter isotopes with $N<90$, is much higher in energy than 
the other members of the band. 
For the same reason as in the case of the quasi-$\beta$ band, 
the experimental bandhead energy of the quasi-$\gamma$ band is 
considerably overestimated.

We have plotted  in Fig.~\ref{fig:nf} the expectation 
value of the $f$-boson number operator 
$\hat{n}_{f}$ obtained with the IBM wave functions of the 
states (a) $0^{+}_{1}$, (b) $2^{+}_{1}$, 
(c) $0^{+}_{2}$, (d) $2^{+}_{2}$, (e) $1^{-}_{1}$, 
and (f) $3^{-}_{1}$. It is remarkable that, 
at $N\approx 88$, both the wave functions of the 
$0^{+}_{1}$ and $2^{+}_{1}$ states 
contain a large amount of $f$-boson components 
$\braket{\hat{n}_{f}}\approx 1$. This suggests that 
the octupole degree of freedom plays an important 
role in the structure of 
the positive-parity ground-state bands at low 
spin for  $N\approx 88$ nuclei. Exception made 
of the Xe 
isotopes, the $0^{+}_{2}$ state appears 
to be of 
double-octupole boson nature because 
$\braket{\hat{n}_{f}}\approx 2$. 
The $f$-boson content of the $2^{+}_{2}$ state 
is similar to that of the $0^{+}_{2}$ state, 
especially for $N\geqslant 88$.
The number of $f$ 
bosons in the wave functions of the 
 $1^{-}_{1}$ 
and $3^{-}_{1}$ states is in the range
$1\leqslant\braket{\hat{n}_{f}}\leqslant 2$.
Note that the contribution of the $f$ boson
to the wave functions is particularly 
large $\braket{\hat{n}_{f}}\approx 2$  
around $N=88$, where the SCMF-PESs 
exhibit the most  pronounced octupole 
deformation effects.

%
%
\begin{figure}[htb!]
\begin{center}
\includegraphics[width=\linewidth]{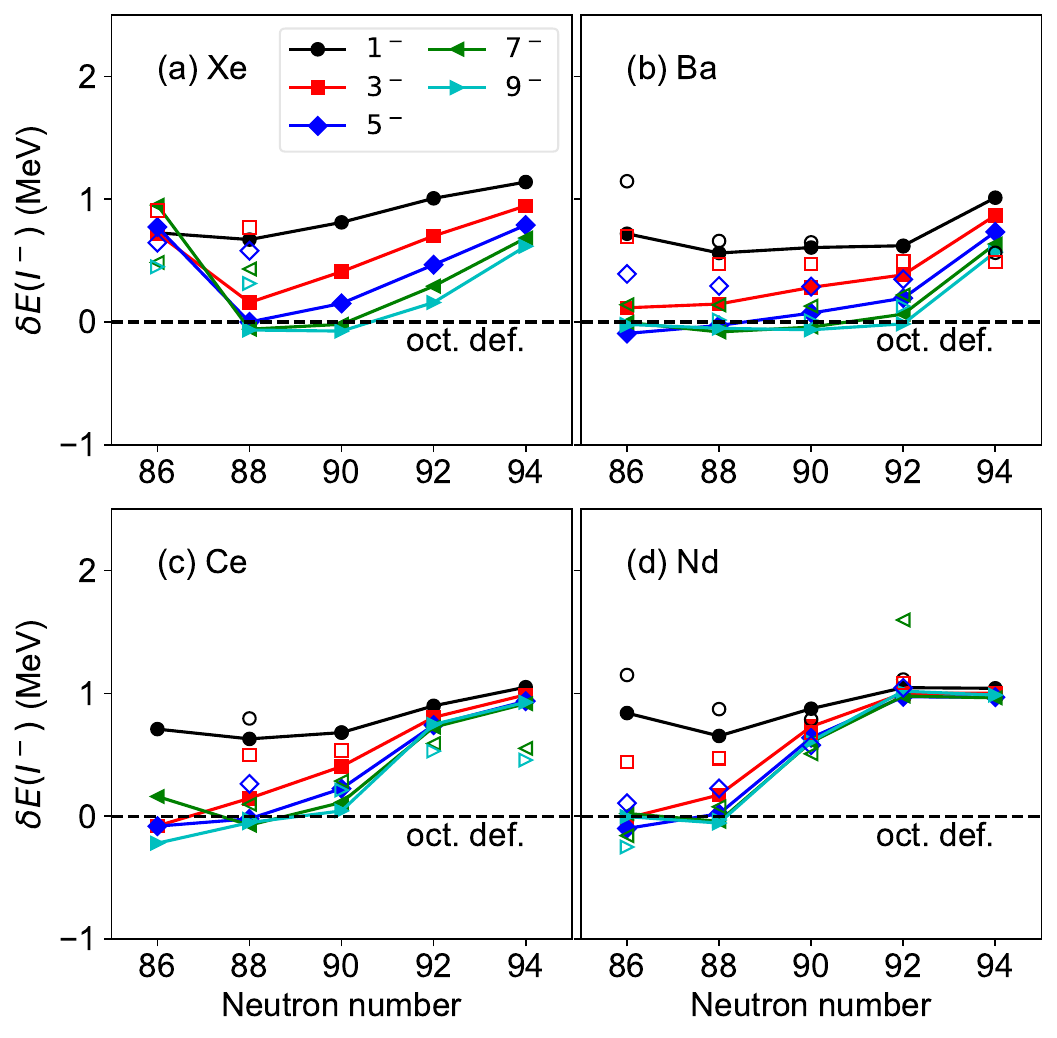}
\caption{The energy displacement $\delta E(I^{-})$ 
(\ref{eq:de}) is plotted 
as a function of the neutron number. 
The theoretical values are connected by lines. Experimental 
 values \cite{data}
for the $I^\pi=1^{-}$, $3^{-}$, 
 $5^{-}$, $7^{-}$, and $9^{-}$ yrast 
states are represented by open
 circles, squares,  diamonds, and left- 
and right-pointing triangles, respectively. 
A broken horizontal line in each panel stands for 
the limit of stable octupole deformation 
$\delta E(I^{-})=0$.}
\label{fig:de}
\end{center}
\end{figure}
%
%
\begin{figure}[htb!]
\begin{center}
\includegraphics[width=\linewidth]{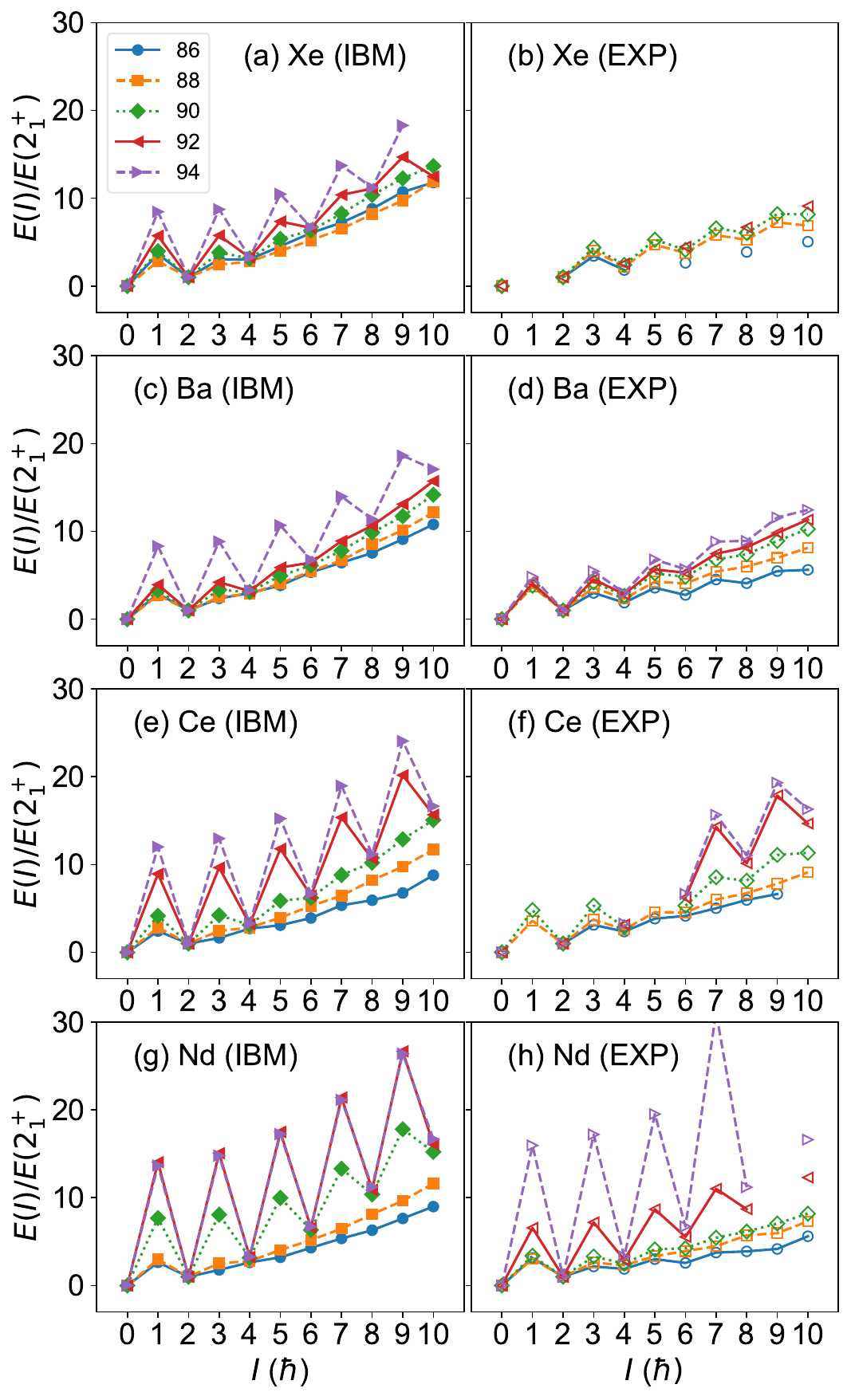}
\caption{The energy ratio 
$E(I^{\pi})/E(2^{+}_{1})$ is depicted as a function of the 
spin $I^{\pi}$. For more details, see the main text.}
\label{fig:alt}
\end{center}
\end{figure}

\subsection{Possible alternating-parity band structure\label{sec:alt}}

In order to 
distinguish whether the members of rotational 
bands are octupole-deformed or octupole
vibrational states, it is convenient to 
analyze the energy displacement defined by 
\begin{align}
\label{eq:de}
\delta E(I^{-}) = E(I^{-}) - \frac{E((I+1)^{+}) + E((I-1)^{+})}{2}, 
\end{align}
where $E(I^{-})$ and $E((I\pm 1)^{+})$ 
represent excitation energies of the odd-spin
negative-parity and even-spin positive-parity yrast states,
respectively. If the positive- and negative-parity bands 
share an octupole deformed bandhead they form an 
alternating-parity doublet and the quantity 
$\delta E(I^{-})$ is equal to zero. The
deviation from the limit $\delta E(I^-)=0$ 
means that the states generating the positive- and negative-parity
bands are very different, and therefore 
the negative parity state must be of  octupole vibrational character.

The results obtained for the energy displacement Eq. (\ref{eq:de}) are 
displayed in Fig.~\ref{fig:de}. For almost all the studied nuclei, the 
energy displacement corresponding to the low-lying negative-parity 
states is close to the limit of stable octupole deformation $\delta 
E(I^-)=0$ for $N\leqslant 90$. However, the $\delta{E}(I^{-})$ values 
obtained for Ce and Nd isotopes with $N>90$ depart sharply from that 
limit. This is in correspondence with the differences observed between 
the low-energy positive- and negative-parity levels in these isotopes 
and the ones observed in the Ba and Xe chains (cf. 
Figs.~\ref{fig:level-pos} and \ref{fig:level-neg}).

As yet another signature of the formation of alternating-parity 
doublets, we study the energy ratio $E(I^{\pi})/E(2^{+}_{1})$. For an 
ideal alternating-parity band, this quantity depends quadratically on 
the spin $I$. On the other hand, in the case of octupole vibrational 
states the positive- and negative-parity bands are decoupled and the 
ratio should also increase quadratically but with a different curvature 
(inverse of the moment of inertia) and therefore a staggering pattern 
as a function of the spin is expected. As can be seen in 
Fig.~\ref{fig:alt} for the Ce and Nd isotopes both the predicted and 
experimental energy ratios $E(I^{\pi})/E(2^{+}_{1})$  increase 
quadratically with  spin $I$ up to $N=88-90$ while a staggering pattern 
emerges for larger neutron numbers. These phenomena are also observed 
in other mass regions, and appear to be characteristic of those nuclei 
where octupole correlations play an important role in low-lying states. 
A similar staggering pattern of the ratio $E(I^{\pi})/E(2^{+}_{1})$ 
appears at $N\approx 134$ in light actinide and at $N\approx 88$ in 
rare-earth regions in the calculations based on the relativistic 
Hartree-Bogoliubov mean field with the DD-PC1 EDF \cite{nomura2014}.

%
%
\begin{figure}[htb!]
\begin{center}
\includegraphics[width=.7\linewidth]{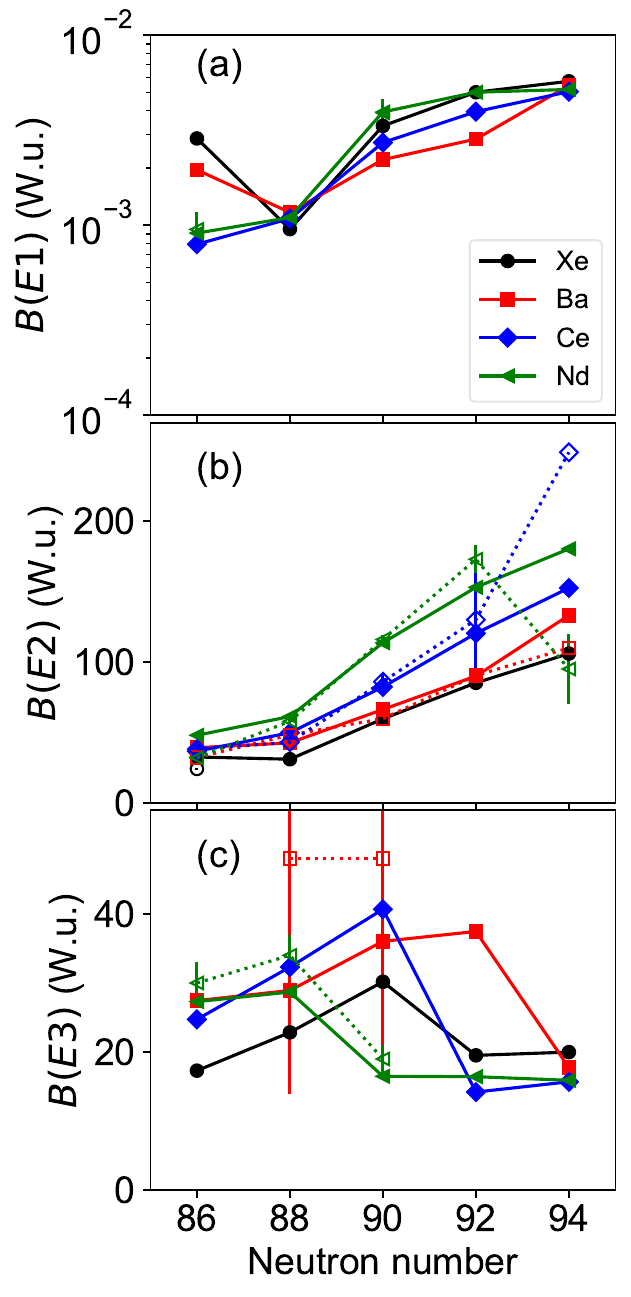}
\caption{
The $B(E1;1^{-}_{1}\to 0^{+}_{1})$ (top), 
$B(E2;2^{+}_{1}\to 0^{+}_{1})$ (middle), and 
$B(E3;3^{-}_{1}\to 0^{+}_{1})$ (bottom) 
reduced transition probabilities (in Weisskopf units)
are compared with the experimental data
\cite{data,bucher2016,bucher2017,kibedi2002}.}
\label{fig:trans}
\end{center}
\end{figure}
%
%
\begin{figure*}[htb!]
\begin{center}
\includegraphics[width=\linewidth]{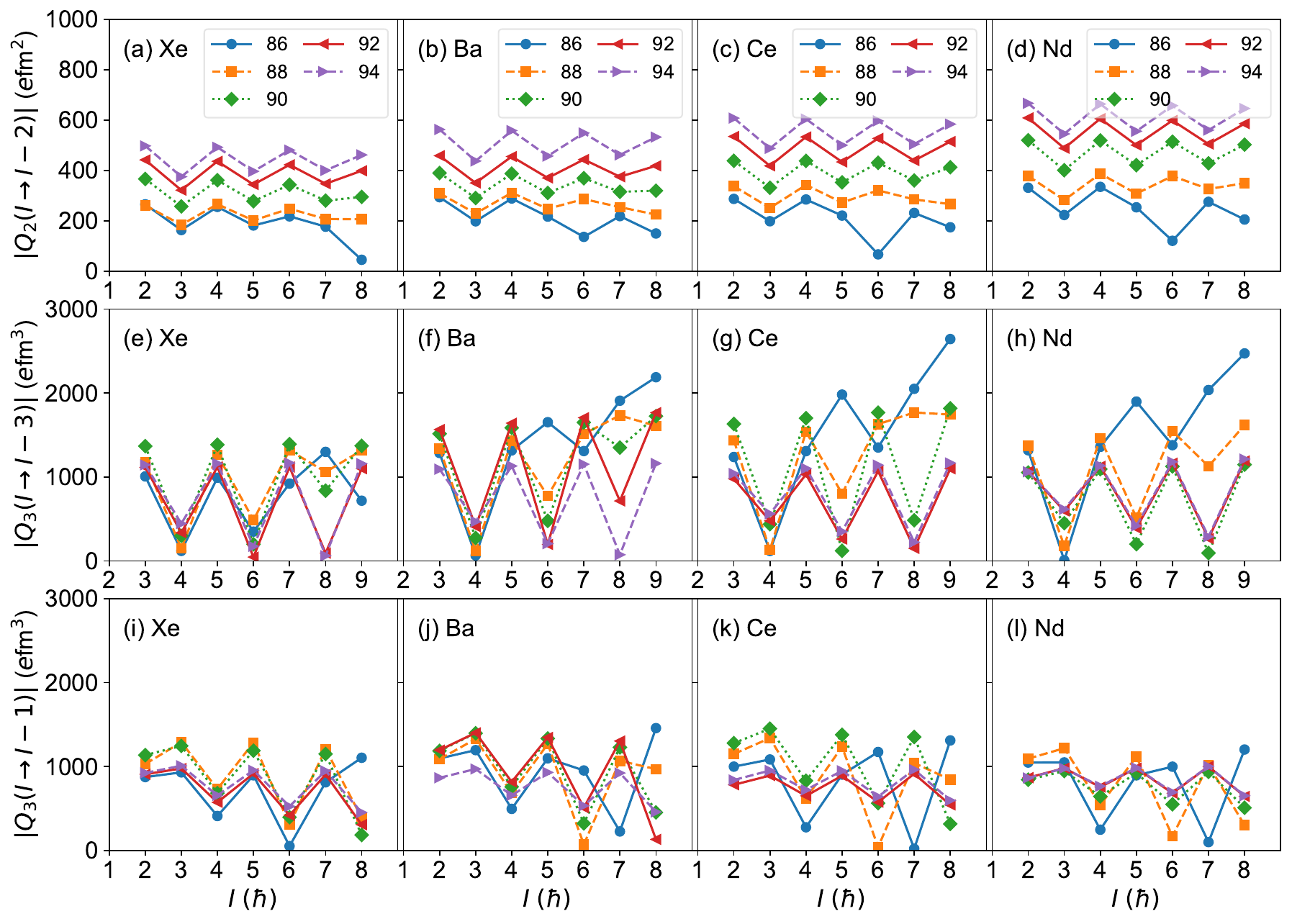}
\caption{The theoretical quadrupole 
$Q_{2}(I\to I-2)$ (top),
octupole  $Q_{3}(I\to I-3)$ (middle), and 
$Q_{3}(I\to I-1)$ (bottom) 
moments in $e$fm$^{\lambda}$ units are plotted as 
as functions of the spin 
$I$. 
Even- (Odd-) $I$ values correspond 
to positive- (negative-) parity states.}
\label{fig:mom}
\end{center}
\end{figure*}

\subsection{Transition rates\label{sec:trans}}

Transition probabilities are computed using the 
electric dipole, quadrupole, and octupole transition 
operators $\hat{T}(E\lambda)$ ($\lambda=1,2,3$)
defined as   
$\hat{T}(E1)=e_{1}(d^{\+}\tilde{f}+f^{\+}\tilde{d})^{(1)}$, 
$\hat{T}(E2)=e_{2}\hat{Q}_{2}$, 
and $\hat{T}(E3)=e_{3}\hat{Q}_{3}$.
 The operators $\hat{Q}_{2}$ and
$\hat{Q}_{3}$ are the same as those introduced 
in Eqs.~(\ref{eq:q2}) and (\ref{eq:q3}), 
respectively. 
The boson effective charges $e_{1}=0.02$ $e$b$^{1/2}$, 
$e_{2}=0.14$ $e$b,  are 
fixed so that the experimental 
$B(E1;1^{-}_{1}\to 0^{+}_{1})$, 
$B(E2;2^{+}_{1}\to 0^{+}_{1})$ transition rates 
are reproduced reasonably well. 
However, in order to fix  $e_{3}$ it is needed to consider the large $B(E3;3^{-}_{1}\to 0^{+}_{1})$  values
observed experimentally in those nuclei where octupole correlations 
are enhanced. In order to account for this fact, the effective $E3$ charge is assumed
to depend on the deformation parameters as 
$e_{3}=0.12\times(1+\bar{\beta_{2}}\bar{\beta_{3}})$ $e$b$^{3/2}$. 
The  $e_{2}$ and $e_{3}$ charges employed in the 
calculations are shown in Figs.~\ref{fig:parameter}(k) 
and \ref{fig:parameter}(l) as functions 
of the neutron number. The behavior of the charge $e_{3}$ 
corresponds to an inverted  
parabola with a maximum around $N=88$, i.e., the 
neutron number for which the global minimum 
of the SCMF-PESs is reflection asymmetric.

The $B(E1;1^{-}_{1}\to 0^{+}_{1})$, 
$B(E2;2^{+}_{1}\to 0^{+}_{1})$, and 
$B(E3;3^{-}_{1}\to 0^{+}_{1})$ reduced 
transition probabilities
obtained in the calculations are compared
with the experimental data 
\cite{data,bucher2016,bucher2017,kibedi2002}
in Fig.~\ref{fig:trans}. The 
$B(E1)$ and 
$B(E2)$ rates agree well with the experimental
ones. They show a steep increase beyond 
$N=88-90$ that is consistent with the development of strong 
collectivity. Moreover, for each isotopic chain, the largest
$B(E3)$ rate corresponds to $N \approx 88$. The computed
$B(E3)$ values  also agree reasonably well with the 
experimental ones. In particular, for $^{144,146}$Ba
the predicted $B(E3)$ rates are within the 
experimental
error bars \cite{bucher2016,bucher2017}. 

We have obtained the transition 
quadrupole $Q_{2}(I\to I')$ and 
octupole $Q_{3}(I\to I')$ moments
from the reduced $E2$ and $E3$ 
matrix elements resulting 
from the diagonalization of the 
$sdf$-IBM Hamiltonian.
Those transition multipole moments 
$Q_{\lambda}(I\to I')$ ($\lambda=2,3$) read
\begin{align}
\label{eq:mom}
 Q_{\lambda}(I\to I')=
&\braket{I'\|\hat{T}(E\lambda)\|I}
\nonumber \\
&\times\sqrt{\frac{16\pi}{(2\lambda+1)(2I+1)}}
(I\lambda 00|I'0)^{-1},
\end{align}
where the factor $(I\lambda 00|I'0)$ 
represents a Clebsch-Gordan coefficient. 

The quadrupole 
$Q_{2}(I^{\pm}\to (I-2)^{\pm})$ and octupole moments 
$Q_{3}(I^{\pm}\to (I-3)^{\mp})$ and 
$Q_{3}(I^{\mp}\to (I-1)^{\pm})$ ($e$fm$^{\lambda}$ units)
are depicted in Fig.~\ref{fig:mom} as functions 
of the spin.
The transitions between  even-spin (odd-spin)
positive (negative) parity states are used to 
compute the quadrupole moment $Q_{2}$ while 
the octupole moments, involve  $E3$ transitions 
between even-spin positive-parity and odd-spin 
negative-parity states.

The $Q_{2}$ moments in each of the in-band 
$I^{\pm}\to (I-2)^{\pm}$ transitions are 
depicted in the top row of
Fig.~\ref{fig:mom} as functions of the spin.
They exhibit a slight odd-even spin staggering pattern. 
Exception are the $Q_{2}$ moments for 
the lightest $N=86$ isotopes, where a certain irregularity 
arises for the spin $I\geqslant6$ because of the change 
in the structure of the wave function around $I=6^{+}$. 
For example, in the case of 
$^{144}$Ce, the states in  the ground-state band
mainly consist of monopole $s$ 
and quadrupole $d$ bosons up 
to $I=4^{+}$. However, for larger 
spin values the 
$f$ boson degree of freedom plays a role 
since $\braket{\hat{n}_{f}}\approx 2$. 
This leads to the small 
$\braket{4^{+} \|\hat{T}(E2)\| 6^{+}}$ 
matrix element. 
In addition, at a given spin $I$, 
the $Q_{2}$ value increases as a function of the 
neutron number, reflecting the increasing 
quadrupole collectivity as the number of 
valence nucleons increases.

The $Q_{3}(I\to I-3)$ moments 
are plotted in 
the middle row of Fig.~\ref{fig:mom}
as functions of the spin. They exhibit a 
staggering pattern. In most of the 
studied isotopic chains, the calculated 
$Q_{3}(I^{+}\to(I-3)^{-})$ moments 
nearly vanish for even spin $I$. 
Here, one should keep in mind that the 
measured reduced $E3$ 
matrix elements between the states 
$I^{+}$ and $(I-3)^{-}$ (with $I$ even) in $^{148,150}$Nd 
\cite{ibbotson1993,ibbotson1997} are close to zero.
Furthermore, the values in the range $1000-2000$ $e$fm$^{3}$
obtained for $Q_{3}(I^{-}\to (I-3)^{+})$
(with $I$ odd) are consistent with the 
experimental data \cite{butler2020b}.

A similar staggering pattern is observed 
for $Q_{3}(I\to I-1)$, plotted 
in the bottom row of Fig.~\ref{fig:mom}
as functions of the spin. In good 
agreement with the empirical trend in this 
mass region 
\cite{butler2016,butler2020b}, the 
$Q_{3}(I^{+}\to(I-1)^{-})$ moments (with 
$I$ even) for $^{148}$Nd decreases with the spin. 
The $Q_{3}(I^{-}\to(I-1)^{+})$ moments (with 
$I$ odd) for the same nucleus are calculated to 
be $\approx 1000$ $e$fm$^{3}$, while experimentally 
the corresponding values are close to zero.

%
%
\begin{figure}[htb!]
\begin{center}
\includegraphics[width=\linewidth]{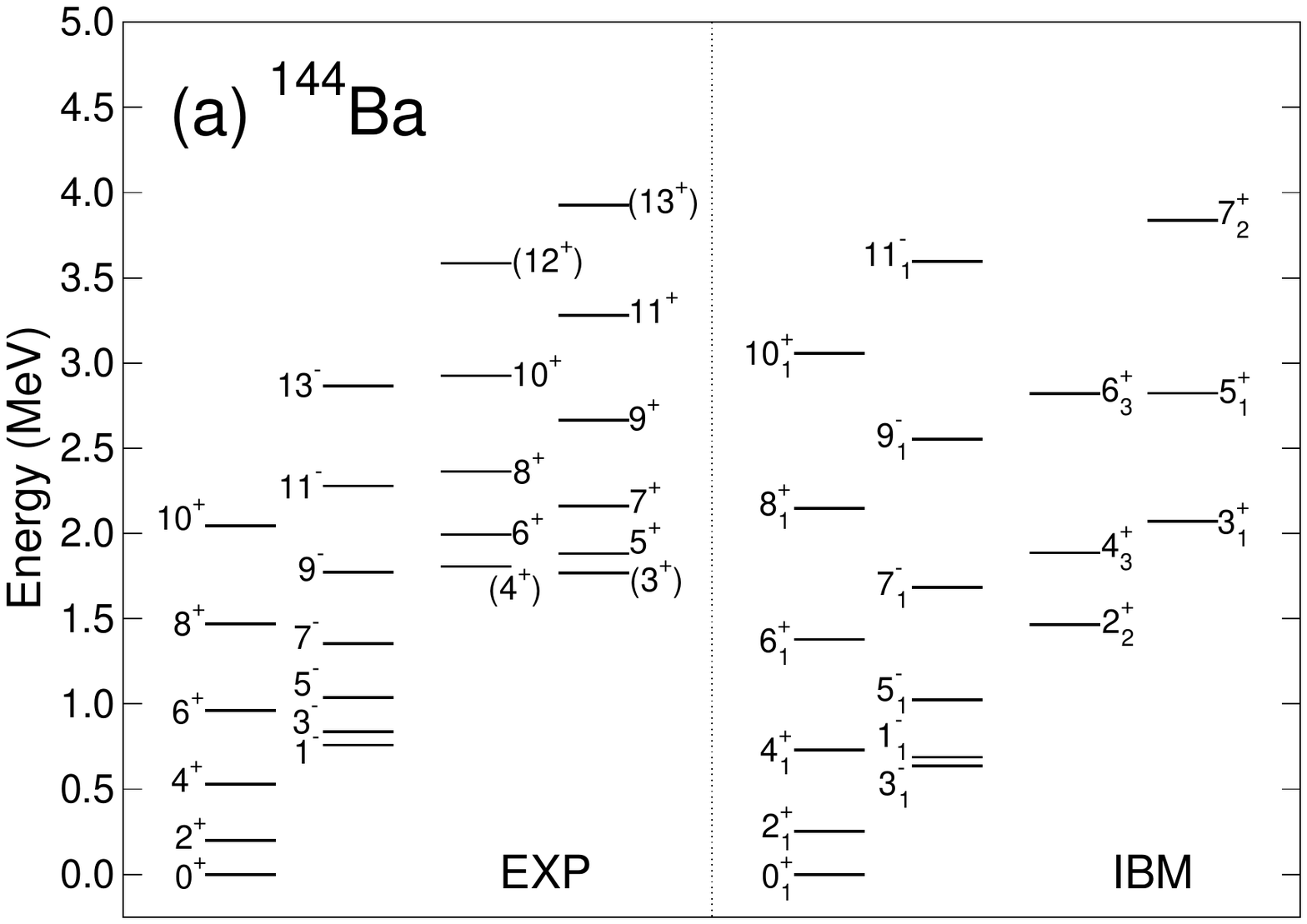}\\
\includegraphics[width=\linewidth]{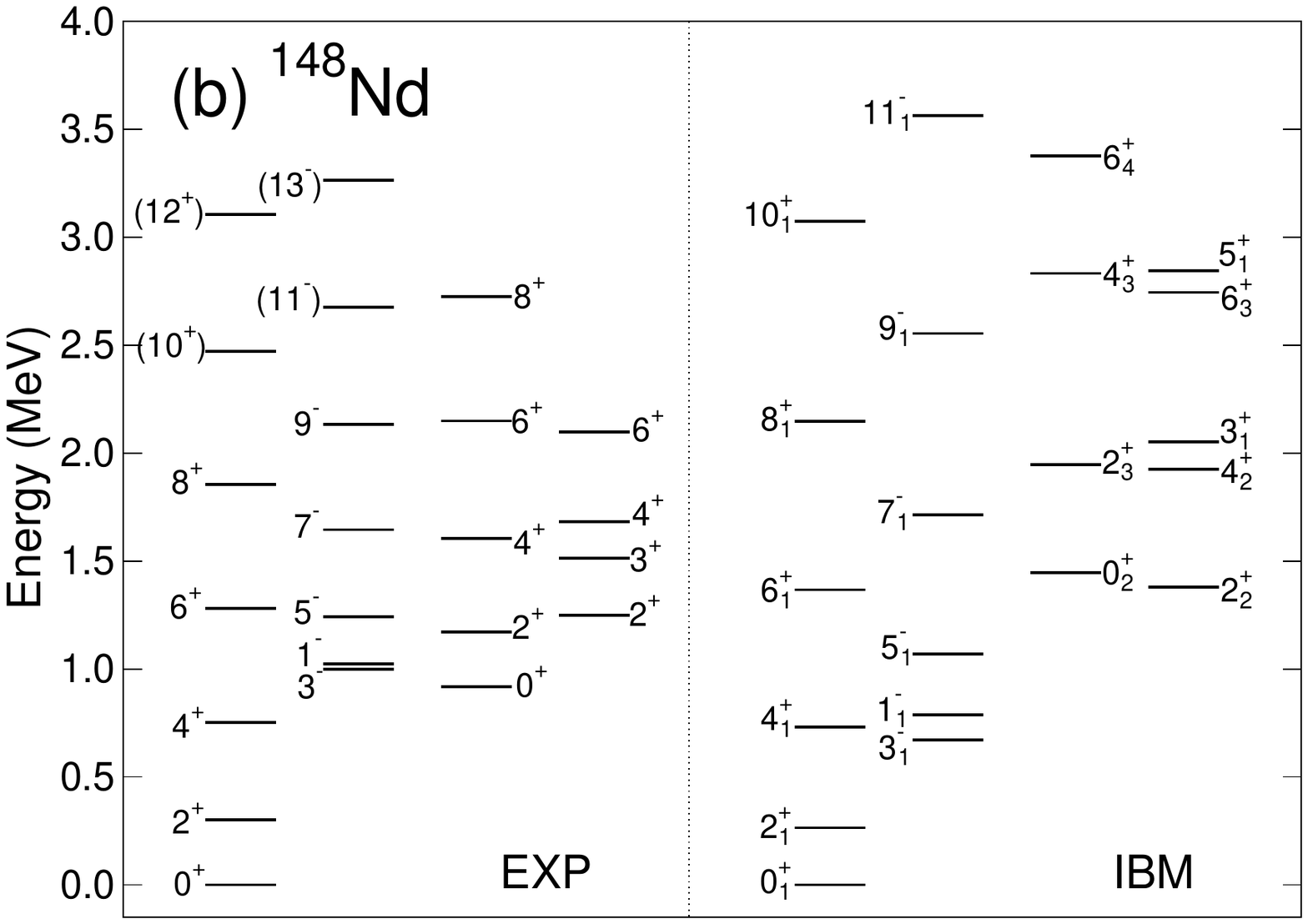}\\
\includegraphics[width=\linewidth]{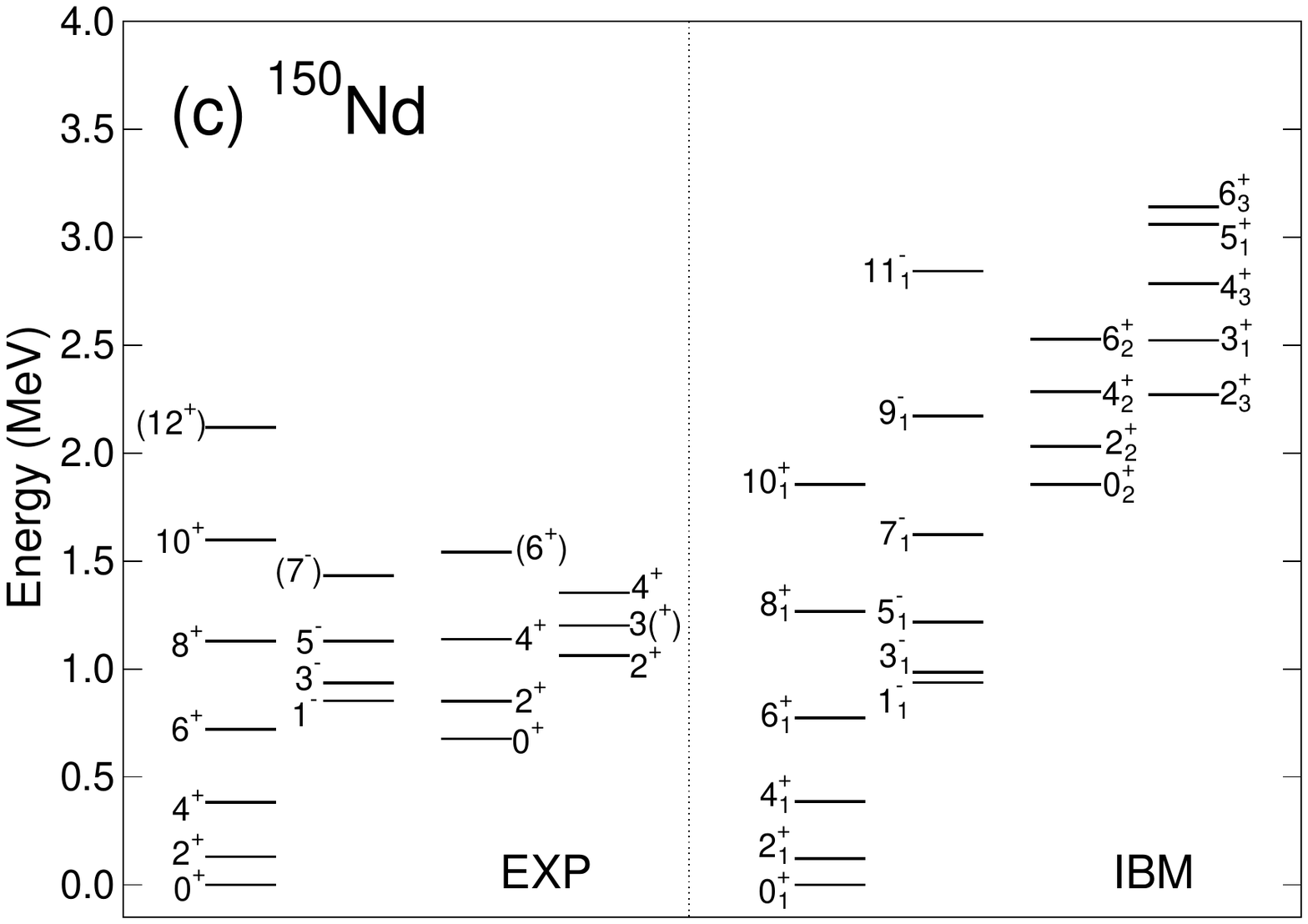}
\caption{Comparison of the theoretical 
and experimental low-energy level
 schemes of (a) $^{144}$Ba, (b) $^{148}$Nd, 
and (c) $^{150}$Nd. The experimental data 
have been  taken from Ref.~\cite{data}.}
\label{fig:level}
\end{center}
\end{figure}

\begin{table}[!htb]
\begin{center}
\caption{\label{tab:em-ba} 
The $B(E1)$, $B(E2)$, and $B(E3)$ transition rates
(in Weisskopf units) obtained for $^{144}$Ba and $^{146}$Ba
are compared with the experimental data 
\cite{bucher2016,bucher2017}.}
 \begin{ruledtabular}
 \begin{tabular}{lccccc}
 & $E\lambda$  & $I_{i}$ & $I_{f}$ & EXP & IBM \\ 
\hline
$^{144}$Ba 
& $E2$ 
  & $2^+_1$ & $0^+_1$ & 48$^{+2}_{-2}$ & 43 \\
& & $4^+_1$ & $2^+_1$ & 86$^{+10}_{-7}$ & 61 \\
& & $6^+_1$ & $4^+_1$ & 54$^{+7}_{-6}$ & 57 \\ 
& & $8^+_1$ & $6^+_1$ & 55$^{+19}_{-12}$ & 37 \\ 
& $E3$ 
  & $3^{-}_1$ & $0^+_1$ & 48$^{+25}_{-34}$ & 29 \\ 
& & $5^{-}_1$ & $2^+_1$ & $<103$ & 50 \\ 
& & $7^{-}_1$ & $4^+_1$ & $<135$ & 63 \\ 
\hline
$^{146}$Ba 
& $E1$ 
  & $1^{-}_1$ & $0^+_1$ & 9.3$^{+0.8}_{-0.7}\times 10^{-7}$ & 2.2$\times 10^{-3}$ \\
& &           & $2^+_1$ & $(6.6\pm0.5)\times 10^{-6}$ & 3.3$\times 10^{-5}$ \\
& & $3^-_1$ & $4^+_1$ & $(1.59\pm0.09)\times 10^{-5}$ & 8.4$\times 10^{-3}$  \\ 
& &         & $2^+_1$ & $(1.84\pm0.13)\times 10^{-6}$ & 2.7$\times 10^{-4}$ \\ 
& $E2$ 
  & $2^+_1$ & $0^+_1$ & 60$\pm 2$ & 66 \\
& & $4^+_1$ & $2^+_1$ & 94$\pm 24$ & 93 \\
& & $6^+_1$ & $4^+_1$ & 93$^{+23}_{-27}$ & 94 \\ 
& & $8^+_1$ & $6^+_1$ & 61$^{+48}_{-24}$ & 73 \\ 
& & $3^-_1$ & $1^-_1$ & 45$\pm 38$ & 50 \\ 
& $E3$ 
  & $3^{-}_1$ & $0^+_1$ & 48$^{+25}_{-29}$ & 36 \\
& & $5^{-}_1$ & $2^+_1$ & 73$^{+88}_{-29}$ & 60 \\
& & $7^{-}_1$ & $4^+_1$ & 82$^{+112}_{-45}$ & 73 \\
& & $9^{-}_1$ & $6^+_1$ & 94$^{+100}_{-94}$ & 39 \\
 \end{tabular}
 \end{ruledtabular}
\end{center} 
\end{table}
\begin{table}[!htb]
\begin{center}
\caption{\label{tab:em-nd}
Same as in Table~\ref{tab:em-ba}, 
but for $^{148}$Nd and $^{150}$Nd. 
Experimental data have been taken from Ref.~\cite{data}.}
 \begin{ruledtabular}
 \begin{tabular}{lccccc}
 & $E\lambda$  & $I_{i}$ & $I_{f}$ & EXP & IBM \\ 
\hline
$^{148}$Nd
& $E1$ 
  & $5^-_1$ & $4^+_1$ & 0.00205$\pm0.00021$ & 0.014 \\ 
& & $7^-_1$ & $6^+_1$ & 0.0043$\pm0.0010$ & 0.020 \\ 
& & $8^+_1$ & $7^-_1$ & 0.0049$\pm0.0011$ & 0.0094 \\ 
& $E2$ 
  & $2^+_1$ & $0^+_1$ & 57.9$\pm 2.2$ & 61 \\
& & $4^+_1$ & $2^+_1$ & 94$\pm 4$ & 92 \\
& & $0^+_2$ & $2^+_1$ & 31.2$\pm 2.2$ & 52 \\
& & $2^+_2$ & $0^+_1$ & 0.54$\pm 0.08$ & 2.6 \\
& &         & $2^+_1$ & 14.4$\pm 1.9$ & 6.0 \\
& &         & $4^+_1$ & 16$\pm 8$ & 4.4 \\
& & $2^+_3$ & $0^+_1$ & 1.9$\pm 0.4$ & 2.6 \\
& & $6^+_1$ & $4^+_1$ & 102$\pm 7$ & 96 \\ 
& & $8^+_1$ & $6^+_1$ & 98$\pm 17$ & 86 \\ 
& & $7^-_1$ & $5^-_1$ & $(1.5\pm0.6)\times 10^{2}$ & 74 \\ 
& $E3$ 
  & $3^{-}_1$ & $0^+_1$ & 34$\pm 3$ & 29 \\
\hline
$^{150}$Nd
& $E1$ 
  & $1^-_1$ & $0^+_1$ & $3.9^{+0.6}_{-0.7}\times10^{-3}$ & $3.9\times10^{-3}$ \\ 
& &         & $2^+_1$ & $7.4^{+0.6}_{-0.7}\times10^{-3}$ & $9.7\times10^{-4}$ \\ 
& & $3^-_1$ & $2^+_1$ & $4.2^{+0.6}_{-0.7}\times10^{-3}$ & $1.1\times10^{-2}$ \\ 
& &         & $4^+_1$ & $4.5^{+0.6}_{-0.7}\times10^{-3}$ & $1.0\times10^{-4}$ \\ 
& & $5^-_1$ & $4^+_1$ & $7^{+9}_{-5}\times10^{-3}$ & $1.8\times10^{-2}$ \\ 
& &         & $6^+_1$ & $7^{+9}_{-5}\times10^{-3}$ & $3.2\times10^{-7}$ \\
& & $2^-_1$ & $2^+_2$ & $4.9^{+2.4}_{-2.0}\times 10^{-5}$ & $1.3\times10^{-2}$ \\
& &         & $2^+_3$ & $(6\pm3)\times10^{-3}$ & 3.5$\times 10^{-4}$ \\
& &         & $3^+_1$ & $3.1^{+1.5}_{-1.2}\times10^{-3}$ & $1.1\times10^{-2}$ \\
& & $3^-_2$ & $2^+_1$ & 8.0$^{+2.0}_{-1.9}\times 10^{-5}$ & $1.0\times10^{-4}$ \\ 
& &         & $2^+_2$ & 8.1$^{+2.0}_{-1.9}\times 10^{-5}$ & $1.4\times10^{-3}$ \\
& &         & $2^+_3$ & $(2.0\pm0.5)\times10^{-3}$ & 7.6$\times 10^{-6}$ \\
& &         & $3^+_1$ & $2.6^{+0.7}_{-0.6}\times10^{-3}$ & 6.4$\times 10^{-5}$ \\
& &         & $4^+_1$ & $(1.7\pm0.4)\times10^{-4}$ & $1.1\times10^{-3}$ \\
& &         & $4^+_2$ & $(3.7\pm0.9)\times 10^{-4}$ & $6.5\times10^{-3}$ \\
& & $7^-_1$ & $6^+_1$ & $(4.3\pm1.0)\times10^{-3}$ & $2.6\times10^{-2}$ \\ 
& & $8^+_1$ & $7^-_1$ & $(4.9\pm1.1)\times10^{-3}$ & 1.8$\times 10^{-5}$ \\ 
& $E2$ 
  & $2^+_1$ & $0^+_1$ & 116$\pm 3$ & 113 \\
& & $4^+_1$ & $2^+_1$ & 180.7$\pm 1.6$ & 162 \\
& & $0^+_2$ & $2^+_1$ & 43.1$\pm 2.3$ & 26 \\
& & $2^+_2$ & $0^+_1$ & 0.7$\pm 0.5$ & 3.1 \\
& &         & $0^+_2$ & $(1.6\pm1.3)\times 10^{2}$ & 42 \\
& &         & $2^+_1$ & 10$\pm 3$ & 4.8 \\
& &         & $4^+_1$ & 19$\pm 7$ & 11 \\
& & $2^+_3$ & $0^+_1$ & 3.0$\pm 0.6$ & 0.11 \\
& &         & $2^+_1$ & $>2.9$ & 0.47 \\
& &         & $4^+_1$ & 1.7$\pm 1.2$ & 1.9  \\
& & $6^+_1$ & $4^+_1$ & 206$\pm 9$ & 175 \\ 
& & $8^+_1$ & $6^+_1$ & 216$\pm 23$ & 175 \\ 
& & $4^+_2$ & $2^+_1$ & 0.015$\pm 0.004$ & 3.3 \\
& &         & $2^+_2$ & 23$\pm 8$ & 59 \\
& &         & $6^+_1$ & 9.2$\pm 2.2$ & 6.5 \\
& & $4^+_3$ & $2^+_1$ & 0.58$\pm 0.20$ & 0.00017 \\
& &         & $2^+_3$ & $(1.3\pm0.5)\times 10^{2}$ & 31 \\
& & $10^+_1$ & $8^+_1$ & 201$\pm 11$ & 165 \\ 
& $E3$ 
  & $3^{-}_1$ & $0^+_1$ & 19$\pm 2$ & 16 \\
 \end{tabular}
 \end{ruledtabular}
\end{center} 
\end{table}

\subsection{Detailed level schemes\label{sec:detail}}

For some of the considered neutron-rich lanthanide nuclei, a wealth of 
experimental data is available regarding the band structure as well as 
the electromagnetic transition rates. In this study, we have examined 
in detail the  low-energy spectra of  $^{144}$Ba, $^{148}$Nd, and 
$^{150}$Nd. The first two of them correspond to $N=88$, for which a 
reflection asymmetric HFB minimum has been obtained. On the other hand, 
$^{150}$Nd exhibits a (reflection symmetric) quadrupole deformed ground 
state with $\beta_{2}\approx0.3$. The theoretical bands presented in 
what follows have been  arranged according to the dominant in-band $E2$ 
transitions. 

Figure~\ref{fig:level}(a) shows the low-energy positive-parity 
bands and the negative-parity band built on the 
$1^{-}_{1}$ state of $^{144}$Ba. 
The experimental data for the non-yrast positive-parity bands 
built on the $4^{+}$ and $3^{+}$ states are also available \cite{zhu2020}, 
and the corresponding theoretical bands are shown in the figure. 
For the yrast bands with both parities, the excitation energies obtained 
for $I\leqslant 5$ agree reasonably well with the experimental data of 
Ref.~ \cite{bucher2016}. However, both bands are somewhat stretched for 
higher spin as compared to the experimental ones. Note, that the 
inversion of the theoretical $1^{-}$ and $3^{-}$ levels might indicate 
that a dipole boson should be included in the calculations, in order to 
lower the $1^{-}$ level. 
The theoretical non-yrast positive-parity bands are much more 
stretched than the observed ones, even though the energies of the 
bandhead are reproduced rather well. 
As discussed earlier, especially for the transitional nuclei around 
$N=88$, due to the substantial degree of configuration mixing 
the repulsion between the low-spin states is so strong as to make 
the description of the band structure difficult particularly for 
non-yrast states.
Another potential explanation is the large number of $f$ bosons present
in the wave functions of the non-yrast states, consequence of the 
large number of them present in our calculation. 

Experimentally, the nucleus $^{146}$Ba also 
exhibits  a stable octupole deformation \cite{bucher2017}. The level 
scheme obtained for $^{146}$Ba is strikingly similar to that of 
$^{144}$Ba. As can be seen from Table~\ref{tab:em-ba}, the $B(E1)$, 
$B(E2)$, and $B(E3)$ reduced transition probabilities obtained for 
$^{144,146}$Ba,  compare well with the experimental values 
\cite{data,bucher2016,bucher2017}. Only the $B(E1)$ rates of $^{146}$Ba 
differ by several orders of magnitude from the experimental ones 
\cite{data}. One should, however, keep in mind that the $E1$ properties 
are mostly determined by non-collective single particle degrees of 
freedom. As the $sdf$-IBM model  is built on collective nucleon pairs 
it is not expected to provide reliable predictions on $E1$ transition 
properties.

The low-energy bands obtained for the isotopes $^{148}$Nd and 
$^{150}$Nd are depicted in Figs.~\ref{fig:level}(b) and 
\ref{fig:level}(c), respectively. 
Empirically, the $^{148}$Nd nucleus is also considered 
\cite{sugawara2007} a transitional nucleus close to the X(5) 
critical point symmetry \cite{iachello2001}. 
For $^{148}$Nd, the agreement with 
the experiment is better than for $^{144}$Ba. This is because the 
number of bosons for the former is large enough so as to provide a 
better quantitative description of the excitation spectra. The  
$1^{-}$, $3^{-}$, and $5^{-}$ energy levels are somewhat lower than
the experimental ones. This correlates well with the especially pronounced 
reflection asymmetric minimum observed in the Gogny-D1M SCMF-PES (cf. 
Fig.~\ref{fig:pesdft}).

Experimental data for $\beta$ and $\gamma$ bands, built on the 
$0^{+}_{2}$ and $2^{+}_{2}$ states, respectively, are available 
for $^{148}$Nd. As can be seen from Fig.~\ref{fig:level}(b), 
the predicted excitation energies of both bands look somewhat irregular 
as compared to the experimental ones. This discrepancy can be attributed 
to a too strong configuration mixing between the low-spin members 
of these bands. 
The excitation energies corresponding to the even-$I$ members of the 
$\gamma$-band  agree reasonably well with the experimental values. 
However, this is not the case for the odd-$I$ members where the 
obtained excitation energies are too high. To improve the description 
of the non-yrast band within the (mapped) IBM calculation, certain 
extensions of the model would be needed. For instance, the inclusion of a 
specific three-body boson interaction in the IBM Hamiltonian lowers the 
energies of odd-$I$ members of the $\gamma$ band \cite{nomura2012tri}

As can be seen in  Fig.~\ref{fig:level}(c), the lowest positive- and 
negative-parity bands of $^{150}$Nd are reasonably well described 
within the mapped IBM calculations. This nucleus was considered 
in Ref.~\cite{lee2018} to be close to the SU(3) limit of the IBM. 
In fact, several features characteristic of the SU(3) dynamical 
symmetry can be observed:  
the SCMF-PES exhibits a large quadrupole (prolate) deformation 
in comparison with the neighboring nucleus $^{148}$Nd; 
the resultant yrast bands for each parity are much more compressed 
and resemble rotational bands (Fig.~\ref{fig:level-pos}(d)); 
the moments of inertia of the positive-parity 
ground-state, quasi-$\beta$, and quasi-$\gamma$ bands are 
approximately equal to each other. 
On the other hand, the energies of 
the $\beta$ and $\gamma$ bandheads are considerably overestimated 
although the energy splittings between the members of the bands are 
well reproduced. The SCMF-to-IBM mapping procedure often yields 
$0^{+}_{2}$ excitation energies somewhat higher than the empirical 
ones. As already pointed out, this mainly comes from the large 
strength parameter $\kappa_{2}$ of the 
quadruple-quadrupole interaction. In some particular deformed nuclei, 
the SCMF-PES suggests a too steep valley around the minimum. To 
reproduce such a topology with the IBM-PES, a large 
quadrupole-quadrupole boson interaction strength is often required. 
To improve the description of the non-yrast states, specifically 
the excitation energies of the excited $0^{+}$ states, some other 
building blocks may need to be included in the mapped $sdf$-IBM 
framework. For instance, the dynamical pairing degree of freedom 
has been introduced in the mapped IBM framework 
\cite{nomura2020pv,nomura2021pv} as additional collective coordinate, 
and have been shown to play a crucial role in lowering the 
$0^{+}_{2}$ energy levels. 

The mapping procedure is also able to provide the energies 
corresponding to non-yrast negative-parity bands. For example, in the 
case of $^{150}$Nd the excitation energies obtained for the $3^{-}_{2}$ 
and $2^{-}_{1}$ states are 2.004 keV and 1678 keV which should be 
compared to the experimental values of  1484 keV and 1435 keV, 
respectively.

Finally, the $B(E1)$, $B(E2)$, and $B(E3)$ transition rates are 
compared with the available experimental data \cite{data} in 
Table~\ref{tab:em-nd}. Exception made of some of the $E1$ rates, the 
overall agreement with the experimental values is reasonable. As discussed
above, the $E1$ strength has a strong dependence on non-collective single
particle degrees of freedom that is impossible to reproduce within the IBM
scheme.

%

\section{Summary\label{sec:summary}}

In this paper, we have examined the onset of octupole deformation and 
the related spectroscopic properties in neutron-rich lanthanide nuclei 
with neutron numbers $86\leqslant N\leqslant 94$ using a mapped IBM 
Hamiltonian obtained from microscopic 
$(\beta_{2},\beta_{3})$-constrained HFB calculations based on the 
Gogny-D1M parametrization. At the mean-field level  
pronounced reflection asymmetric global minima emerge around $N=88$. 
For larger neutron numbers reflection symmetric ground states are 
obtained and the corresponding quadrupole deformations increase with 
neutron number.

Spectroscopic properties have been studied via the diagonalization of  
the $sdf$-IBM Hamiltonian, with the strength parameters 
obtained by mapping the SCMF-PES onto the expectation value 
of the Hamiltonian in the boson condensate state. 
The patterns exhibited by the predicted  
low-energy negative-parity spectra and $B(E3;3^{-}_{1}\to 0^{+}_{1})$ 
transition rates point towards enhanced octupolarity 
(Fig.~\ref{fig:level-neg}) in  $N \approx 88$ isotopes. In particular, 
the results obtained for the energy displacement 
$\delta E(I)$ (Fig.~\ref{fig:de}) and ratio $E(I)/E(2^{+}_{1})$ 
(Fig.~\ref{fig:alt}) indicate the occurrence of an approximate 
alternating-parity doublet structure around $N=88$. For $N\geqslant 90$ 
separate positive and negative bands are obtained and an octupole 
vibrational regime characteristic of a $\beta_{3}$-soft potential 
develops.

The results obtained in this work for neutron-rich lanthanide  and the 
ones already obtained for rare-earth \cite{nomura2015} and light 
actinide \cite{nomura2020oct,nomura2021oct} nuclei indicate that the 
octupole degree of freedom plays an important role to describe the 
structural evolution  and spectroscopic properties of the low-lying 
states in certain regions of the nuclear chart. In particular, the 
employed Gogny-EDF-based $sdf$-IBM framework provides a reasonable 
description of key spectroscopic properties that help to identify the 
interplay between the quadrupole and octupole degrees of freedom in the 
lanthanide region. The results  of the present analysis encourage us to 
explore the relevance of octupole deformations in other mass regions. 
Within this context, proton-rich nuclei with $Z\approx N\approx 56$ 
appear as plausible candidates to be considered in future work. Work 
along these lines is in progress and will be reported in a forthcoming 
article. 

On the other hand, disagreements between the mapped IBM results 
and spectroscopic data indicate needs for assessing the quality of 
the mapping, that is, whether the problems lie on the fermionic 
calculations or the mapped boson Hamiltonian. 
There are several prescriptions along this direction: 
(i) For Xe and Ba isotopes, the 
calculation overestimated the $6^{+}_{1}$ and $8^{+}_{1}$ 
excitation energies (see Fig.~\ref{fig:level-pos}). 
In this region, neutrons (protons) 
occupy $f_{7/2}$ ($g_{7/2}$) orbitals, giving rise to 
the $J=6^{+}$ nucleon pairs. 
The inclusion of the corresponding $i$ boson in the IBM may lower 
the energies of those yrast states with $I\geqslant6^{+}$. 
(ii) As seen from Fig.~\ref{fig:level-neg}, the mapped IBM was not 
able to reproduce in detail the systematic of the $7^{-}$ energy 
in Ba and Nd isotopes. It would be worthwhile to see if 
the fermionic calculation within the GCM can reproduce 
the empirical tendency of this state. 
(iii) The calculated quasi-$\beta$ (Fig.~\ref{fig:level-beta}) and 
quasi-$\gamma$ bands (Fig.~\ref{fig:level-beta}) turned out to be 
considerably higher than the experimental ones. As we discussed in 
Sec.~\ref{sec:spectra}, this is partly due to the too strong 
quadrupole-quadrupole interaction strength. 
In the deformed region, it is also well known that hexadecapole 
($g$) boson plays an important role. 
It would be then interesting to reveal whether the inclusion of the 
$g$ bosons in the mapping procedure, either explicitly or by using 
the renormalization method of Ref.~\cite{otsuka1985}, can improve 
the description of the non-yrast states. 
(iv) The conventional mapping procedure within the generalized seniority 
scheme of the nuclear shell model \cite{OAI,mizusaki1996} has been 
successful for those nuclei near the closed shells. The results obtained 
from such a method for nearly spherical and vibrational nuclei could 
be compared with those from the SCMF-to-IBM mapping procedure.

\begin{acknowledgments}
This work has been supported by the Tenure Track Pilot Programme of the 
Croatian Science Foundation and the \'Ecole Polytechnique F\'ed\'erale 
de Lausanne, and the Project TTP-2018-07-3554 Exotic Nuclear Structure 
and Dynamics, with funds of the Croatian-Swiss Research Programme. The  
work of LMR was supported by Spanish Ministry of Economy and 
Competitiveness (MINECO) Grant No. PGC2018-094583-B-I00. The work of JEGR has 
been partially supported by the Ministerio de Ciencia e Innovaci\'on 
(Spain) under projects number PID2019-104002GB-C21, by the 
Consejer\'{\i}a de Econom\'{\i}a, Conocimiento, Empresas y Universidad 
de la Junta de Andaluc\'{\i}a (Spain) under Group FQM-370, by the 
European Regional Development Fund (ERDF), ref.\ SOMM17/6105/UGR, and 
by the European Commission, ref.\ H2020-INFRAIA-2014-2015 (ENSAR2). 
Resources supporting this work were provided by the CEAFMC and the 
Universidad de Huelva High Performance Computer (HPC@UHU) funded by 
ERDF/MINECO project UNHU-15CE-2848. 
\end{acknowledgments}

\bibliography{refs}

\end{document}